\pdfoutput=1

\documentclass[aps,twocolumn,showpacs,amssymb,amsmath,nofootinbib,floatfix,superscriptaddress,showkeys]{revtex4-1}

\usepackage{graphicx,color}
\usepackage{amsmath,amssymb,bm}
\usepackage{float}
\usepackage{latexsym}
\usepackage{comment}
\usepackage{hyperref}
\usepackage{epsfig}
\usepackage{makecell}
\usepackage[caption=false]{subfig}
\begin{document}

\title{Hollowness in $pp$ and $p \bar p$ scattering in a Regge model}

\author{Wojciech Broniowski}
\email{Wojciech.Broniowski@ifj.edu.pl}
\affiliation{The H. Niewodnicza\'nski Institute of Nuclear Physics, Polish Academy of Sciences, 31-342 Cracow, Poland}
\affiliation{Institute of Physics, Jan Kochanowski University, 25-406 Kielce, Poland}

\author{L\'aszl\'o Jenkovszky}
\email{jenk@bitp.kiev.ua}
\affiliation{Bogolyubov Institute for Theoretical Physics (BITP),
Ukrainian National Academy of Sciences \\14-b Metrologicheskaya str., Kiev, 03680, Ukraine}

\author{Enrique Ruiz Arriola}
\email{earriola@ugr.es}
\affiliation{Departamento de F\'isica At\'omica, Molecular y Nuclear and Instituto Carlos I de
  Fisica Te\'orica y Computacional,  Universidad de Granada, E-18071  Granada, Spain}

\author{Istv\'an Szanyi}
\email{sz.istvan03@gmail.com}
\affiliation{Uzhgorod National University, 14 Universytets'ka str., Uzhgorod, 88000, Ukraine}

\begin{abstract}

The proton-proton and proton-antiproton inelasticity profiles
in the impact parameter display very interesting and sensitive features
which cannot be deduced solely from the current large body of high-energy scattering data. In particular,
phenomenological studies exhibit a link
between the ratio of the real to imaginary parts of the elastic scattering amplitude at a finite momentum transfer,
and the corresponding change of character of the inelastic processes from central to peripheral collisions.  
We describe how a theoretical model, accommodating the existing data, based on the
Regge hypothesis including both the Pomeron and odderon as double poles,
and $\omega$ and $f$ mesons as single poles in the complex-$J$ plane, generates a hollow in the
inelasticity at low impact parameters. The hollowness effect, which generally may be sensitive to model details, 
does unequivocally take place both for $pp$ and $p \bar p$ collisions within the
applied Regge framework, indicating inapplicability of inelasticity-folding geometric approaches.

\end{abstract}

\date{12 June 2018}



\maketitle

\section{Introduction \label{sec:intro}}

Scattering experiments with hadrons are usually designed to learn
about their structure and interactions~\cite{Barone:2002cv}. In the
case of proton-proton ($pp$) and proton-antiproton ($p \bar p$) collisions, a wealth
of differential elastic scattering data has been collected since the
mid 1950's above center-of-mass (CM) energies of $\sqrt{s}=6$ GeV, characterized by elastic
diffractive scattering. Accordingly, the data exhibit a peak at soft kinematics,
i.e., at small momentum transfers $ -t \ll s$ (for recent comprehensive reviews of the data and models see,
e.g..,~\cite{Dremin:2012ke,Pancheri:2016yel}).
Despite the abundant experimental information and numerous theoretical efforts, it is fair to say that 
we lack a truly working approach based directly on the fundamental
Quantum Chromodynamics (QCD) in the non-perturbative 
soft regime $-t \lesssim \Lambda_{\rm QCD}^2 \ll s$. 
This situation has stimulated
the use of pre-QCD ideas and models which embody not only the desirable
theoretical constraints such as unitarity, crossing, and analyticity,
but also display the outstanding experimental features of the data. These models and the 
following parameterizations
have been steadily and quantitatively tested and improved along the
years. Regge theory, while phenomenological and not fundamental from the
QCD viewpoint, satisfies these important theoretical constraints and at the same time
is flexible enough as to allow for a uniform quantitative description of the data.

A complementary and enlightening way of visualizing the high-energy
scattering results is by passing, via the  Fourier-Bessel transform, from the momentum transfer $t$ 
to the impact parameter  $b$. This variable is conjugate to $\sqrt{-t}$, with 
$b \sim 1/\sqrt{-t}$.  In 1963 van
Hove introduced the inelasticity profile or the overlap function~\cite{van1963phenomenological,VanHove:1964rp} (see
also~\cite{Amaldi:1979kd} and the references therein), which corresponds
to the impact parameter distribution of the inelastic cross section.
This representation has a transparent interpretation, since different impact
parameters decouple from one another. A major issue in this regard is
the fact that the inelastic profile depends on the phase of the scattering
amplitude and thus is not determined solely from the differential
elastic scattering cross section {\em without} some additional assumptions. 
Despite this generic source of arbitrariness, most $pp$ analyses in the wide range of 
10~${\rm GeV} \le \sqrt{s} \le$~500~${\rm GeV}$ have provided a shape for
the inelasticity profile which is compatible with a natural expectation that the most
inelastic collisions are central, i.e., the inelasticity profiles have a maximum at $b=0$.

Whereas this central maximum of inelasticity is in fact explicitly
implemented in geometric
models~\cite{Chou:1968bc,Chou:1968bg,Cheng:1987ga,Bourrely:1978da,Block:2006hy,Block:2015sea}
which are quite naturally based on folding of partonic distributions in the impact
parameter space, there is no particular {\it a priori } reason why it
should be so. In fact, recent
papers~\cite{Alkin:2014rfa,Dremin:2014eva,Dremin:2014spa,%
Arriola:2016bxa,RuizArriola:2016ihz,Albacete:2016pmp,Dremin:2016ugi,Broniowski:2017aaf,%
Dremin:2017ylm,Anisovich:2014wha,Troshin:2016frs,Troshin:2017zmg,Troshin:2017ucy}
have reported a mounting evidence
suggesting that this paradigm may change in the light of the
measurements by the TOTEM collaboration at the CERN Large Hadron Collider (LHC). The found 
{\em hollowness} feature shows that the
inelasticity profile becomes maximal at a finite value of $b$, whereas at $b=0$ it has a local minimum. 

In this paper we reanalyze this issue in a simple Regge
model which, as will be shown, provides efficiently a reasonable description of
a large $pp$ and $p\bar p$ elastic scattering data in the range
10~${\rm GeV} \le \sqrt{s} \le$~13~${\rm TeV}$. The main advantage of the
Regge framework is that it not only predicts the $s$ dependence at
small $t$, but also fixes the total amplitude, and hence a
fit to the elastic differential cross section allows one to determine both the
modulus and the argument of the amplitude. Of course, there
are many Regge models on the market, so the question of
uniqueness of the description is a pertinent one; we leave a
thorough comparison of different model proposals and parameterizations for a future research
and here focus on a particularly simple Regge model.

\section{The basics \label{sec:stat}}

In this section we introduce our notation and methodology in a way that our problem
can be easily stated, and motivate in passing our use of the Regge theory
within this context. In general, the $NN$ and $N \bar N$ elastic scattering amplitudes
have 5 independent complex components which can only be determined
from a complete set of experiments involving 9 observables, such as
differential cross sections and polarization data~\cite{puzikov1957construction}.
The amplitudes fulfill the crossing
relations~\cite{Gribov:1963gx,Sharp:1963zz}.  As is customary in
such studies, we neglect the spin dependence. Whereas the degree of uncertainty introduced by this approximation
is not known,
the spin-flip amplitudes have been found to be non-vanishing but small in
E950 fixed target experiment at $\sqrt{s} \sim 25$~GeV at the BNL Relativistic Heavy-Ion Collider (RHIC) (for
a review see, e.g., \cite{Selyugin:2015vao} and references
therein). Thus a small but systematic error in the amplitude is
foreseen. This is an important observation for the statistical
analysis of the data which allows for a more relaxed interpretation of the
$\chi^2$ minimization than the conventional one, as will be used
in Section~\ref{sec:model}

\subsection{The phase ambiguity}

An important issue which is of relevance in our analysis regards
the uniqueness of the amplitude obtained from the differential cross section
via fitting analysis in the presence of absorption, as it is the case for
$pp$ and $p \bar p$ scattering at $t < 0 $ and $ s > 4 M^2$, with $M$ denoting the proton mass. 
Indeed, if $f(s,t)$ is the elastic scattering amplitude,
\begin{eqnarray}
f(s,t) = {\rm Re} \, f(s,t) + i \, {\rm Im} \, f(s,t) 
\end{eqnarray}
the invariant differential elastic cross section is given
by
\begin{eqnarray}
\frac{d \sigma_{el}}{dt}= \frac{\pi}{p^2} |f(s,t)|^2, \label{eq:diff}
\end{eqnarray}
where $p=\sqrt{s/4-M^2}$ 
denotes the CM momentum of the proton.
Only the absolute value $|f(s,t)|$ enters Eq.~(\ref{eq:diff}), thus 
in the notation
\begin{eqnarray}
f(s,t) = |f(s,t)|\frac{\rho(s,t) + i}{\sqrt{1+\rho(s,t)^2}}
\end{eqnarray}
the real function $\rho(s,t)$, defined as
\begin{eqnarray} 
\rho(s,t) = \frac{{\rm Re} \, f(s,t)}{{\rm Im} \, f(s,t)}
\end{eqnarray}
 remains unconstrained by the elastic scattering
data alone. This freedom merely reflects the incomplete information on
the system. Of course, the $(s,t)$ dependent phase of the scattering
amplitude is not arbitrary and has a physical significance~\cite{Gersten:1969ae} 
(for a review see, e.g., \cite{Bowcock:1976ax}). 

In quantum mechanics, this ambiguity is resolved by analyticity in the
scattering potential at a fixed distance. Likewise, the fixed-$t$ dispersion relations
have been suggested as a possible way to circumvent the ambiguity
problem in $pp$ or $p \bar p$ scattering, since they impose analyticity in the $s$ variable for the
scattering amplitude $f(s,t)$.
Thus, up to subtractions (which may depend on $t$), the real and imaginary
parts are related to each other via the dispersion relation
\begin{eqnarray}
{\rm Re}\, f(s,t) = \frac1{\pi} \int_{4 M^2}^\infty ds' \frac{{\rm Im} \, f(s',t)}{s'-s}.
\label{eq:fixed-t}
\end{eqnarray}
For $t=0$ one has the optical theorem 
\begin{eqnarray}
{\rm Im}\, f(s,0) = \frac{p}{4\pi}\sigma_{\rm tot}(s).
\label{eq:optical}
\end{eqnarray}

If one uses the crossing-odd variable 
\begin{eqnarray}
\nu=\frac{s-u}{4M} = \frac{2s-t-4M^2}{4M},
\end{eqnarray} 
one has the crossing relation $f_{pp} (\nu,t) = f_{p \bar p}
(-\nu,t)^*$~\cite{Gribov:1963gx,Sharp:1963zz} for the central
interaction.  In the limit $s \ll t$, it yields $f_{pp} (s,t) =
f_{p \bar p} (-s,t)^*$, such that one can write a fixed-$t$ dispersion
relation for the odd and even combinations,  $f_\pm (s,t)=
f_{p \bar p} (s,t) \pm f_{pp} (s,t) $, independently. In this limit, and neglecting
the threshold effect which can be done at large $s$, 
the family of functions $\beta(t) s^{\alpha(t)}$ fulfills
Eq.~(\ref{eq:fixed-t}), since for $-1 < \alpha < 0$ one has the identity 
\begin{eqnarray}
\frac1{\pi}\int_{0}^\infty ds' s'^{\alpha} \left[\frac1{s'-s} \pm \frac1{s'+s} \right] 
= \frac{(-s)^\alpha \pm (s)^\alpha}{\sin \pi \alpha}.
\end{eqnarray}
Here we take $-s = |s| e^{i \pi}$ and proceed by analytic
continuation, making the necessary subtractions for other values of
$\alpha$. Therefore, the Regge theory, where the amplitude reads
$f(s,t)=\sum_i \beta_i (t) s^{\alpha_i(t)}$, does indeed satisfy
the fixed-$t$ dispersion relations for $s \ll t$. The Regge amplitude is odd under
crossing $pp \to p \bar p$~\footnote{This in particular refers to
single Regge poles, but it is also valid for derivatives with respect
to $\alpha$ which arise for the $n$-fold Regge poles, see Section~\ref{sec:model}.}. 
These features justify our motivation to take a particular phenomenologically based
realization of a Regge theory in the following sections.

\subsection{Impact parameter and the overlap function \label{sec:imp}}

The Fourier-Bessel transform of the amplitude $f(s,t)$ is denoted as
$p \,h(b,s)$~\cite{Amaldi:1979kd}, with $p$ denoting the CM momentum 
of the proton,
\begin{eqnarray}
2p\, h(b,s)=  2\int_0^\infty \!\!\! q dq J_0(bq) f(s,-q^2). \label{eq:invf}
\end{eqnarray} 
Next, we present a glossary of formulas for the total, elastic, and
inelastic cross sections in the $b$ representation:
\begin{eqnarray} 
\sigma_{\rm tot}(s)  &=& 4 p \int d^2 b \, {\rm Im} \, h(b,s),  \label{eq:st}\\
\sigma_{\rm el}(s) &=& 4 p^2 \int d^2 b\,  |h( b,s)|^2,  \label{eq:sel} \\
\sigma_{\rm in}(s) &\equiv& \sigma_{\rm tot}(s) - \sigma_{\rm el}(s) = \int d^2 b \, \sigma_{\rm in} (b,s). \label{eq:sin}
\end{eqnarray} 
Here, the dimensionless integrands $\sigma_{\rm tot} (b,s)$, $\sigma_{\rm el} (b,s)$, and
$\sigma_{\rm in} (b,s)$  can be
interpreted as profiles representing the $b$-dependent relative number of the appropriate
collisions.  The
inelasticity profile is equal to
\begin{eqnarray}
\sigma_{\rm in} (b,s)  = 4p \, {\rm Im}\,  h(b,s) -  4p^2|h(b,s)|^2.   \label{eq:prof}
\end{eqnarray} 
Unitarity and positivity of absorption imply
\begin{eqnarray}
1 \ge \sigma_{\rm in}(b,s) \ge 0,  \label{eq:pos}
\end{eqnarray}
whereas the condition 
\begin{eqnarray}
\sigma_{\rm in}(b,s) \le 2 (2p \, {\rm Im} \, h(b,s)) - (2p \,{\rm Im} \, h(b,s))^2
\end{eqnarray}
yields, consistently, the upper bound \mbox{$\sigma_{\rm in}(b,s) \le 1$}.  

The criterion for
hollowness is to have a minimum of $\sigma_{\rm in}(b,s)$ at
$b=0$, i.e.,
\begin{eqnarray}
 \left . \frac{d\sigma_{\rm in}(b,s)}{db^2} \right |_{b=0} < 0. \label{eq:crit}
\end{eqnarray}
At this point it should be noted that the phase ambiguity discussed in Section~\ref{sec:stat} is
transferred to the impact parameter space, hence the very issue of
hollowness cannot be decided based just on the elastic scattering data, without
further theoretical or model input. In fact, in~\cite{Broniowski:2017rhz} it
has been shown that adopting various admissible choices of $\rho(s,t)$ influences
quantitatively and qualitatively the 
result. 

\subsection{The exponential fall-off and hollowness}

The most characteristic feature of the high-energy elastic scattering
is the diffraction peak, which is characterized by 
the slope parameter at the origin, defined as
\begin{eqnarray}
B(s) = \left . \frac{d}{dt} \ln \frac{d \sigma_{el}(s,t)}{dt} \right |_{t=0}.
\end{eqnarray}
A simple Gaussian profile in the momentum transfer 
(note an assumed $t$-independent $\rho(s)$ function)
\begin{eqnarray}
f(s,t)= \frac{p \sigma_{\rm tot}(s)}{4\pi}[ i + \rho(s)] e^{B(s)t/2} \label{eq:gau}
\end{eqnarray}
fulfills the optical theorem. In the $-t \ll s $ limit, one has
\begin{eqnarray}
\sigma_{\rm el}(s) = \int_{-s+4 M^2}^0 (-dt) \frac{d \sigma_{\rm el}}{dt}(s,t) \to 
\int_{-\infty}^0 (-dt) \frac{d \sigma_{\rm el}}{dt}(s,t) \nonumber \\
\end{eqnarray}
and the following equation is satisfied:
\begin{eqnarray}
\frac{\sigma_{\rm el}(s)}{\sigma_{\rm tot}(s)}= \frac{[1+\rho(s)^2] \sigma_{\rm tot} (s)}{16 \pi B(s)}. \label{eq:rel}
\end{eqnarray}
This relation has been observed to work quite accurately in a
soft-Pomeron $pp$ and $p \bar p$ model for a fit range of $5~{\rm GeV}
< \sqrt{s} < 500$~GeV~\cite{Covolan:1996uy}. The model applied
in Section~\ref{sec:model} also fulfills relation (\ref{eq:rel}) to an accuracy better
than $5\%$ in the whole fitting range. 

The Fourier-Bessel transform of the
exponential profile (\ref{eq:gau}) is
\begin{eqnarray}
2 p h(b,s) = \frac{\sigma_{\rm tot}(s)}{4\pi B}[ i + \rho(s)] e^{-b^2 / 2B}.
\end{eqnarray}
Substitution of this form into  Eq.~(\ref{eq:prof}) at $b=0$ implies, via
the positivity condition ~(\ref{eq:pos}), that
\begin{eqnarray}
\sigma_{\rm el}(0,s) = \frac{\sigma_{\rm tot}(s)}{4\pi B}[ 2 - 4 \frac{\sigma_{\rm el}(s)}{\sigma_{\rm tot}(s)}] \ge 0, 
\end{eqnarray}
and thus 
\begin{eqnarray}
\sigma_{\rm el}(s) \le {1 \over 2} \sigma_{\rm tot}(s),
\end{eqnarray}
in accordance with experiment.  Thus, in the Gaussian model the {\em
	largest} elastic cross section which can be achieved is half
	of the total cross section, which shows that the scattering in
	this model is intrinsically inelastic.

In order to better appreciate this point, it is worth to consider a
situation for an arbitrary inelasticity profile, however, with a sharp
edge at, say, $b=R$,
\begin{eqnarray}
2 p h(b,s) = 0, \qquad b \ge R .
\end{eqnarray}
Thus, we have  
\begin{eqnarray}
\sigma_{\rm in} &=& \int_0^R 2\pi b db \left[ 4p \,{\rm Im}\, h(b,s) -  4p^2|h(b,s)|^2\right] , \nonumber \\ 
\sigma_{\rm tot}&=& \int_0^R 2\pi b db \left[ 4p \,{\rm Im} \, h(b,s) \right]. 
\end{eqnarray}
Our goal is to {\it maximize} $\sigma_{\rm in} $ for a general
complex profile $h(b,p)$ with a {\it fixed} $\sigma_{\rm tot}$ which
can be readily done by maximizing 
\begin{eqnarray}
\max_{h(p,s)} \left[ \sigma_{\rm in} - \lambda  \sigma_{\rm tot} \right] ,
\end{eqnarray}
with $\lambda$ a Lagrange multiplier. We get from the corresponding
Euler-Lagrange equations  ${\rm Re} h(p,s) =0$, and thus 
\begin{eqnarray}
1- \lambda - 2 p\, {\rm Im} \,h(b,s) =0,
\end{eqnarray}
which implies a {\it constant} profile. For such a situation, the {\it smallest} 
possible elastic cross section is $\sigma_{\rm el}= \sigma_{\rm tot} /2$ with a black-disk 
geometry $\sigma_{\rm in}(b,s)=1$ for $b \le R$, yielding $\sigma_{\rm in}(s)=\pi R^2$. 
Therefore, if $\sigma_{\rm el}(s) < \sigma_{\rm tot}(s) /2$, as
happens experimentally,
the edge cannot be sharp and a gray disk picture sets in.

Turning to the Gaussian profile, the curvature of the inelasticity
profile at the origin is
\begin{eqnarray}
\frac{1}{2} \left . \frac{d^2 \sigma_{\rm in}(b,s)}{db^2} \right |_{b=0} = 
\frac{64 \pi \sigma_{\rm el}^2 (4\sigma_{\rm el}(s)- \sigma_{\rm tot}(s))}{\left(\rho(s)^2+1\right)^2 \sigma_{\rm tot}(s)^4},
\end{eqnarray}
such that the turnover to hollowness takes place at~\cite{Broniowski:2017rhz}
\begin{eqnarray}
\frac{\sigma_{\rm el}(s)}{\sigma_{\rm tot}(s)}={1 \over 4}, \label{eq:to}
\end{eqnarray}
a fact that will also follow to a good accuracy in the more sophisticated Regge model discussed in Section~\ref{sec:model}.

\section{The dipole Regge model \label{sec:model}}

One of the most remarkable successes of the Regge theory was the early
prediction of a diffraction pattern in high energy
collisions~\cite{Barone:2002cv}. However, the conventional Regge theory
based on single Regge poles in the complex-$J$ plane does not account
easily for the dip or the bump structures unveiled in
the CERN Interacting Storage Rings (ISR) experiments~(see, e.g., \cite{Amaldi:1979kd} and references
therein), hence modifications became mandatory. The
Barger and Phillips empirical parameterization~\cite{Phillips:1974vt},
which was successful in fitting the early data and was improved recently by the inclusion of form
factors~\cite{Fagundes:2013aja}, does not provide an energy dependence
stemming from Regge ideas, and hence does not comply to the fixed-$t$
dispersion relations.  

However, as noted many years ago, multiple
Regge poles are not only not forbidden, but may in fact naturally occur
quantum mechanically~\cite{Bell:1964fz,bialkowski1970phenomenology}.  They are actually suggested by
the dual models with the Mandelstam analyticity (see,
e.g., \cite{Jenkovszky:1981xv} for an impact parameter analysis).  The
model with different Regge trajectories, including the double-pole odderon,
was thus proposed~\cite{Jenkovszky:1976sf} and later extended for finite
$t$ \cite{Saleem:1981av} and the odderon
rise~\cite{Jenkovszky:1987gv,Jenkovszky:1990ki}. The 
upgraded version was described in~\cite{Jenkovszky:2011hu}.

Let us mention in this regard that whereas the order of the Regge pole
cannot be fixed by first principles, the Froissart bound prevents poles
or order higher than 3, and the requirement $\sigma_{\rm el} \le \sigma_{\rm tot} $ prevents asymptotically a moving triple
pole~\cite{Martynov:2007dy}. 

In the light of the recent TOTEM measurements at $\sqrt{s}=13$~TeV,
the double pole Pomeron is preferred compared to a single- or
triple-pole Pomeron~\cite{Alkin:2011if}. The statement is based on
dispersion relations for the meson-proton and proton-proton forward
elastic scattering.  Recent data from the TOTEM Collaboration at 13
TeV provide a convincing evidence on the existence of the
odderon~\cite{Martynov:2017zjz}, discarding many of the models on the
market not including this particular element. Further successful fits
were proposed in~\cite{Khoze:2017swe,Khoze:2018bus,Martynov:2018nyb}.

In this paper, we consider the spin-averaged case of the invariant
high-energy scattering amplitudes, which are sums of four
terms \cite{Jenkovszky:2011hu}.  The two asymptotically leading terms
are the Pomeron (P) and the odderon (O), and two secondary
contributions come from the $f$ and $\omega$ Regge poles.

We note that $P$ and $f$ have positive
$C$, thus enter  the scattering amplitude with the
same sign in $pp$ and $p \bar p$ scattering, whereas $O$ and
$\omega$ have negative $C$, thus enter with opposite signs:
\footnote{Here we use the normalization where
${d\sigma_{\rm el}\over{dt}}(s,t)={\pi\over s^2}|A(s,t)|^2$ and $\sigma_{\rm tot}(s)={4\pi\over s}\,{\rm Im} \, A(s,t=0)$.}
\begin{eqnarray}
A\left(s,t\right)_{pp}^{p \bar p}=A_P\left(s,t\right)+A_f\left(s,t\right)
\pm\left[A_{\omega}\left(s,t\right)+A_O\left(s,t\right)\right]. \nonumber \\ \label{Eq:Amplitude}
\end{eqnarray}
The model of  Eq.~(\ref{Eq:Amplitude}) may be
extended by adding more Reggeons, whose role becomes
increasingly important towards lower energies. In our fits at relatively large $s$, their contribution
can be effectively absorbed in $f$ and $\omega$ \cite{KKL1}.

\begin{center}
	\begin{table*}[tb]
		\caption{Optimum values and the uncertainties of the model parameters following from a joint
		fit to $pp$ and $p\bar p$ data for elastic differential cross section, the total cross section, and the parameter~$\rho$. 
		See text for details. \label{tab:parameters}}	
		\begin{ruledtabular}
			\begin{tabular}{cccccccc}
				\multicolumn{1}{c}{}&\multicolumn{2}{c}{Pomeron}& \multicolumn{2}{c}{Odderon}&\multicolumn{2}{c}{Reggeons}&\multicolumn{1}{c}{} \\
				\hline 
				&$a_P~[\sqrt{\rm mb\,GeV^2}]$ & $360$ (fixed)  &$a_O~[\sqrt{\rm mb\,GeV^2}]$&$1.75\pm0.11$& $a_f~[\sqrt{\rm mb\,GeV^2}]$ & $-20.05\pm0.17$& \\
				&$b_P$ & $4.19\pm0.15$  &$b_O$ &$0.914\pm0.007$& $b_f~[{\rm GeV}^{-2}]$  & $0$ (fixed) &         \\
				&$\delta_P$ & $0.0293\pm0.0005$ &$\delta_O$&$0.275\pm0.005$&$\alpha_{0f}$&0.703 (fixed)& \\
				&$\alpha'_{P}~[{\rm GeV}^{-2}]$& $ 0.5069\pm0.0028$  &$\alpha'_{O}~[{\rm GeV}^{-2}]$ &$0.2309\pm0.0017$&$\alpha'_{f}~[{\rm GeV}^{-2}]$&0.84 (fixed)& \\
				&$\varepsilon_P$ &$0.278\pm0.015$&$\varepsilon_O$&$1.318\pm0.003$&$s_{0f}~[{\rm GeV}^2]$&$1$ (fixed)&        \\
				&$s_{0P}~[{\rm GeV}^2]$& $100$ (fixed) &$s_{0O}~[{\rm GeV}^2]$ &$100$ (fixed)&$a_{\omega}~[\sqrt{\rm mb\,GeV^2}]$&$10.65\pm0.64$& \\ \cline{1-5}
				\multicolumn{1}{c}{}&\multicolumn{2}{c}{}&\multicolumn{2}{c}{}&$b_{\omega}~[{\rm GeV}^{-2}]$&$0$ (fixed)&\\ 
				&&&&&$\alpha_{0\omega}$&0.435 (fixed)&\\
				&${\rm NDF}=159$&$\chi^2=223.5$&$\chi^2/{\rm NDF}=1.4$&&$\alpha'_{\omega}~[{\rm GeV}^{-2}]$&0.93 (fixed)&\\
				&&&&&$s_{0\omega}~[{\rm GeV}^2]$&$1$ (fixed)&\\
			\end{tabular}
		\end{ruledtabular}
	\end{table*}
\end{center}

Secondary Reggeons are parametrized in a standard way \cite{KKL, KKL1}, with linear Regge trajectories and exponential residua. 
The $f$ and $\omega$ Reggeons are the principal non-leading contributions to $pp$ or $p \bar p$ scattering:
\begin{equation}\label{Reggeon1}
A_f\left(s,t\right)=a_f{\rm e}^{-i\pi\alpha_f\left(t\right)/2}{\rm e}
^{b_ft}\Bigl(s/s_{0f}\Bigr)^{\alpha_f\left(t\right)},
\end{equation}
\begin{equation}\label{Reggeon2}
A_\omega\left(s,t\right)=ia_\omega{\rm e}^{-i\pi\alpha_\omega\left(t\right)/2}{\rm e}
^{b_\omega t}\Bigl(s/s_{0\omega}\Bigr)^{\alpha_\omega\left(t\right)},
\end{equation}
with $\alpha_f\left(t\right)=\alpha_{0f}+\alpha'_{f}t$ and $\alpha_{\omega}\left(t\right)=\alpha_{0\omega}+\alpha'_{\omega}t$.

\begin{figure}[b]
\begin{center}
\includegraphics[width=0.49\textwidth]{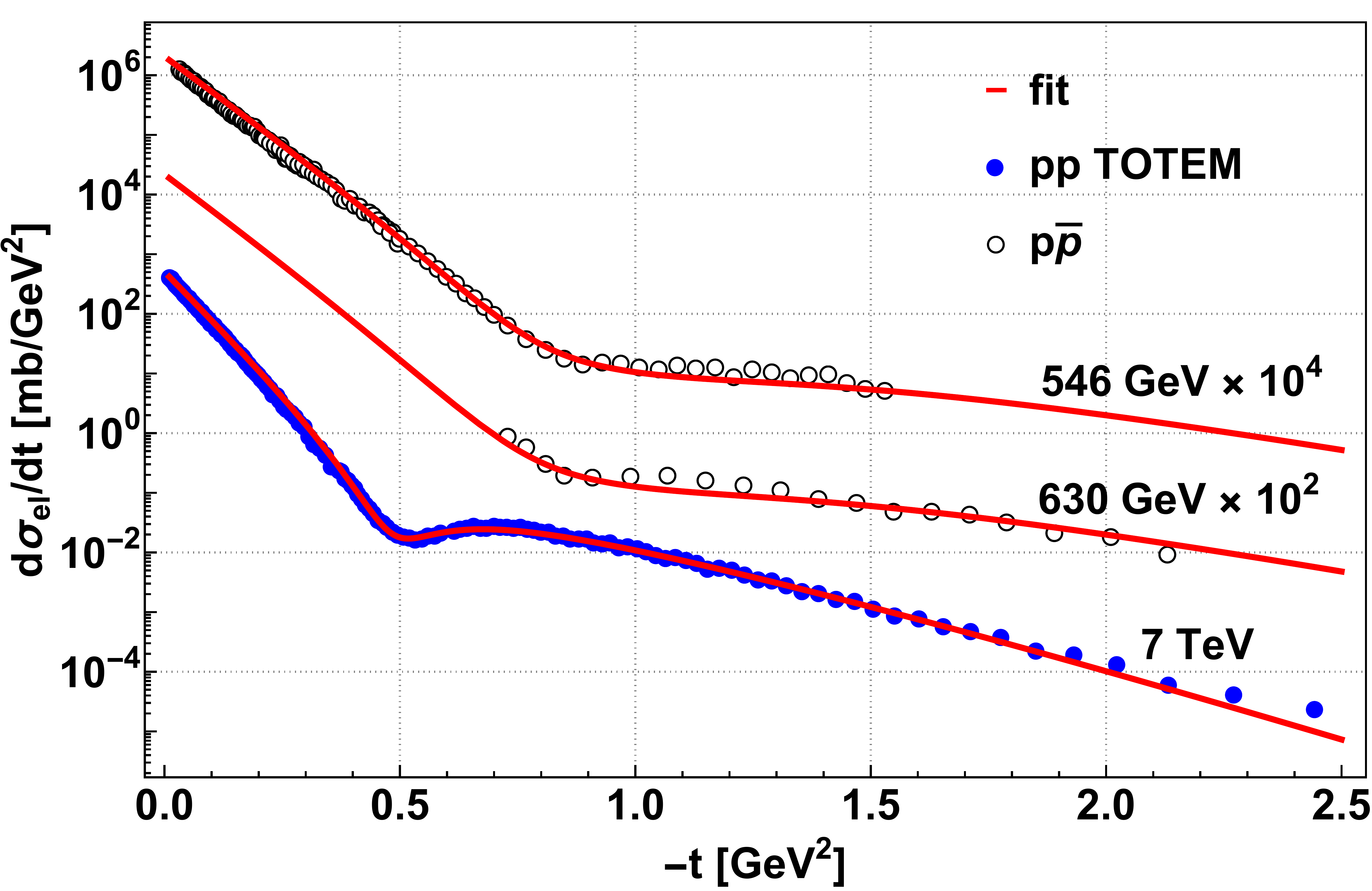}
\end{center}
\vspace{-5mm}
\caption{
Fit to the $pp$ and $p\bar p$ differential elastic cross sections at several collision energies, plotted as a function of the 
momentum transfer $-t$, compared to the data of Refs.~\cite{totem7,barppdcs}.
\label{fig:sel}}
\end{figure}

\begin{figure*}[tb]
\begin{center}
\subfloat[]{%
\includegraphics[width=0.5\textwidth]{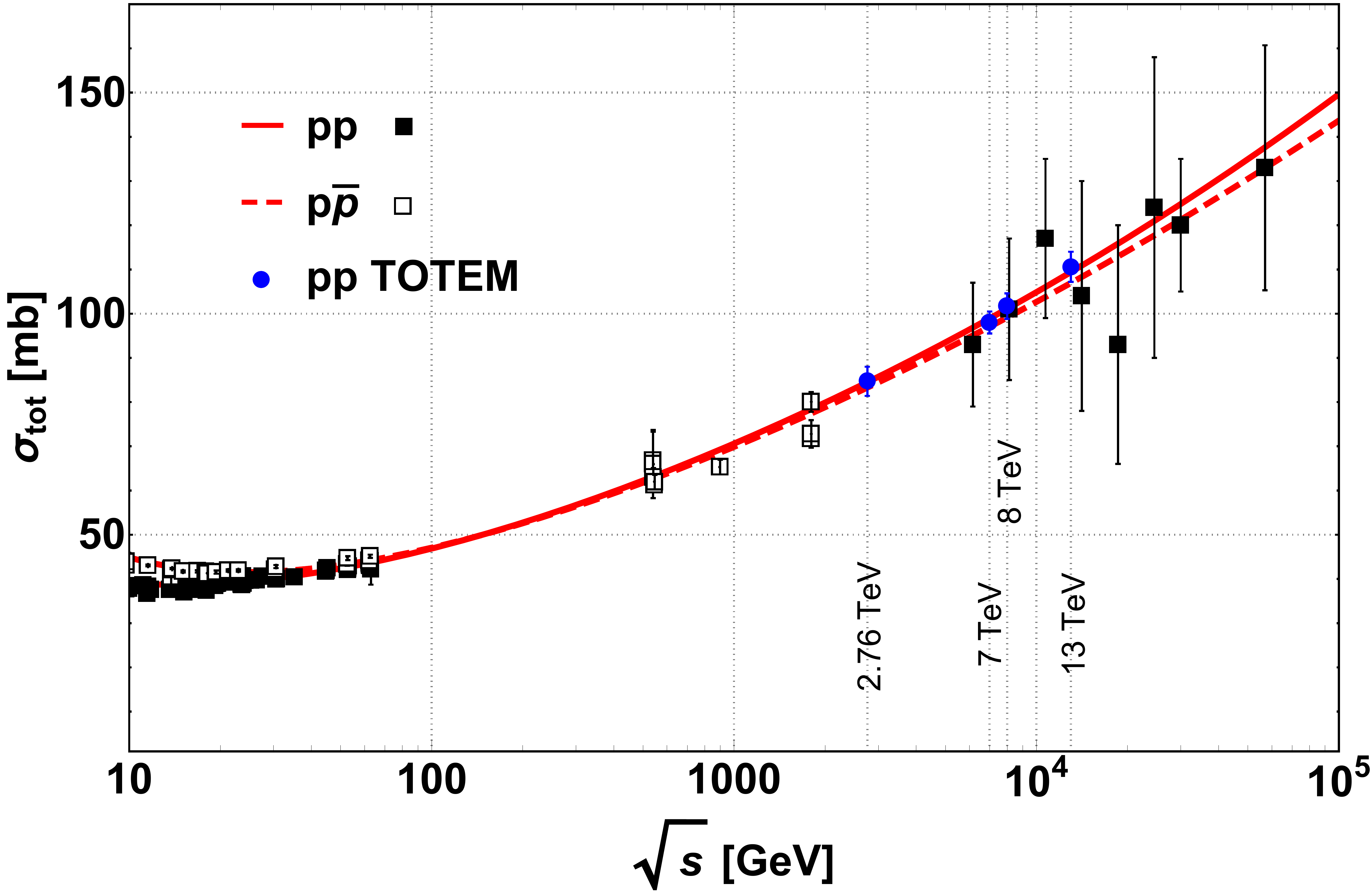}
}
\subfloat[]{%
\includegraphics[width=0.5\textwidth]{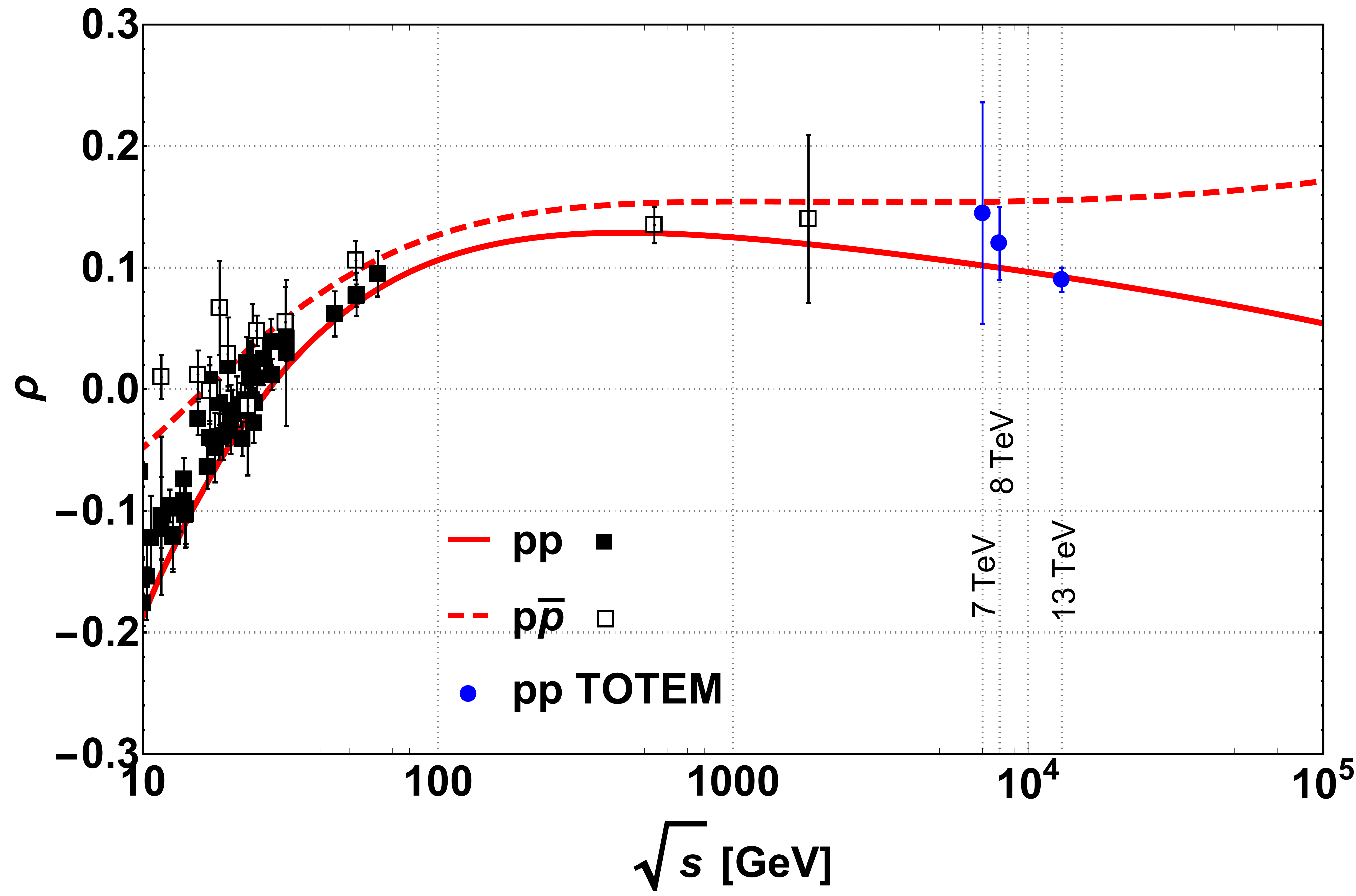}
}
\end{center}
\vspace{-5mm}
\caption{a): Total $pp$ and $p\bar p$ cross sections plotted as a functions of $\sqrt{s}$, 
compared to the data of Refs.~\cite{Auger,totem81,totem72,PDG,Giani}. b): The ratio of the real to imaginary 
part of the elastic $pp$ and $p\bar p$ amplitude at $t=0$, plotted as a function of $\sqrt{s}$ and compared to the data of Refs.~\cite{totem7,totem82,PDG,TOTEM_rho}.
The lines indicate our joint fit.
The LHC collision energies are indicated with the vertical labels.
\label{fig:sigma}}
\end{figure*}

As already mentioned, the Pomeron is a dipole in the $J$-plane
\begin{eqnarray}\label{Pomeron}
& &A_P(s,t)={d\over{d\alpha_P}}\Bigl[{\rm e}^{-i\pi\alpha_P/2}G(\alpha_P)\Bigl(s/s_{0P}\Bigr)^{\alpha_P}\Bigr]= \\ \nonumber
& &{\rm e}^{-i\pi\alpha_P(t)/2}\Bigl(s/s_{0P}\Bigr)^{\alpha_P(t)}\Bigl[G'(\alpha_P)+\Bigl(L_P-i\pi
/2\Bigr)G(\alpha_P)\Bigr].
\end{eqnarray}
Since the first term in squared brackets determines the shape of the cone, one fixes
\begin{equation} \label{residue} G'(\alpha_P)=-a_P{\rm
	e}^{b_P[\alpha_P-1]},
\end{equation} 
where $G(\alpha_P)$ is recovered
by integration. Consequently, the Pomeron amplitude of Eq.~(\ref{Pomeron}) may be rewritten in the following ``geometric'' form 
(for details see \cite{PEPAN} and references therein):
\begin{eqnarray}\label{GP}
A_P(s,t)&=&i{a_P\ s\over{b_P\ s_{0P}}}[r_{1P}^2(s){\rm e}^{r^2_{1P}(s)[\alpha_P-1]}\\ \nonumber
& &-\varepsilon_P r_{2P}^2(s){\rm e}^{r^2_{2P}(s)[\alpha_P-1]}],
\end{eqnarray} 
where $r_{1P}^2(s)=b_P+L_P-i\pi/2$, $r_{2P}^2(s)=L_P-i\pi/2$, \mbox{$L_P\equiv\ln(s/s_{0P})$}, and the Pomeron trajectory is
\begin{eqnarray}\label{Ptray}
\alpha_P\equiv \alpha_P(t)&=& 1+\delta_P+\alpha'_{P}t.
\end{eqnarray}

The odderon contribution (labeled with the subscript ``$O$'') is assumed to be of the same form as for the Pomeron, apart for different values of the adjustable parameters:
\begin{eqnarray}\label{Odd}
A_O(s,t)&=&{a_O\ s\over{b_O\ s_{0O}}}[r_{1O}^2(s){\rm e}^{r^2_{1O}(s)[\alpha_O-1]}\\ \nonumber 
& &-\varepsilon_O r_{2O}^2(s){\rm e}^{r^2_{2O}(s)[\alpha_O-1]}],
\end{eqnarray}
where $r_{1O}^2(s)=b_O+L_O-i\pi/2$, $r_{2O}^2(s)=L_O-i\pi/2$, \mbox{$L_O\equiv\ln(s/s_{0O})$}, and the trajectory is
\begin{eqnarray}\label{Eq:Otray}
\alpha_O\equiv \alpha_O(t) &= &1+\delta_O+\alpha'_{O}t.
\end{eqnarray}

The free parameters of the model were simultaneously fitted to the data on the differential elastic $pp$ and $p\bar p$ 
cross section, as well as to the data on the total cross section and the ratio
\begin{equation}
\rho(s)=\frac{{\rm Re}\,  A(s,t=0)}{{\rm Im} \, A(s,t=0)}.
\end{equation}
The fit was done by using the MIGRAD algorithm of MINUIT~2 with the data in the following intervals:
\begin{itemize}
\item for $pp$ differential elastic cross section at 7 TeV \cite{totem7}: $0.35~{\rm GeV^2}\leqslant-t\leqslant2.5~{\rm GeV^2}$;
\item for $p\bar p$ differential elastic cross section at 546 and 630 GeV \cite{barppdcs}: $0.5~{\rm GeV^2}\leqslant-t\leqslant2.5~{\rm GeV^2}$;
\item for $pp$ and $p\bar p$ total cross section and parameter $\rho$: $20~{\rm GeV}\leqslant\sqrt{s}\leqslant57$~TeV 
\cite{totem7,Auger,totem81,totem72,totem82,PDG,Giani,TOTEM_rho}.
\end{itemize}

\begin{figure}[tbh]
	\begin{center}
		\includegraphics[width=0.49\textwidth]{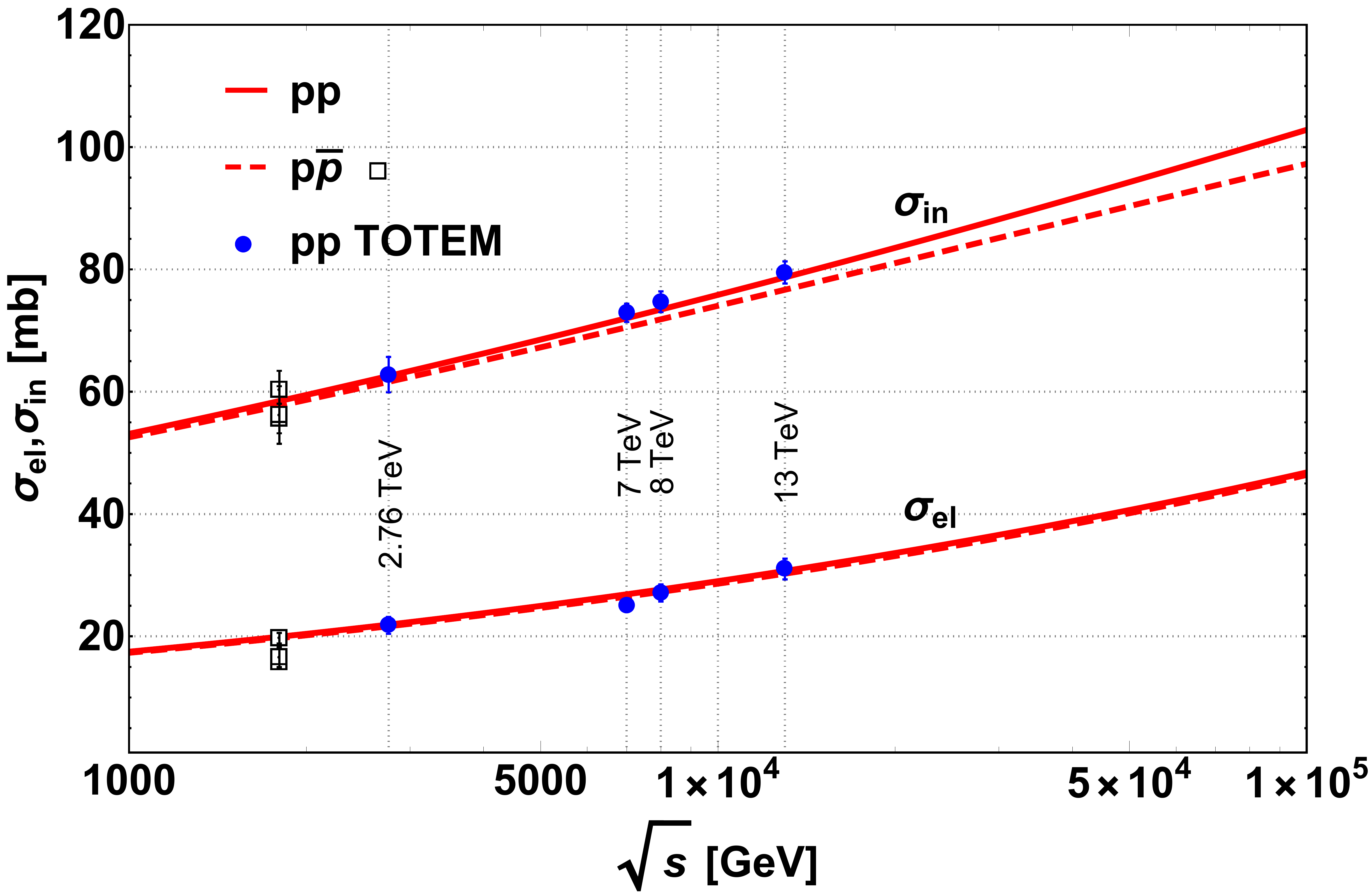}
	\end{center}
	\vspace{-5mm}
	\caption{Elastic and inelastic $pp$ and $p \bar p$ cross sections, plotted as functions of $\sqrt{s}$
	and compared to the data of Refs.~\cite{totem81,totem72,PDG,Giani}. The lines indicate our fit.
		\label{fig:sigma_el_in}}
\end{figure}

The above intervals in $t$ were chosen to optimize our fit, whereby the number of the outliers in the differential cross section was reduced to 63.
The optimum values of the fitted parameters and the values of $\chi^2$ 
are collected in Table~\ref{tab:parameters}. Figures \ref{fig:sel}-\ref{fig:sigma} show the quality of the model
fit to the world data. We note that the overall agreement catches all the features of the data. The value of $\chi^2/{\rm NDF}=1.4$ 
indicates a need for improvement on the theoretical model side, as remarked in Section~\ref{sec:stat}.

The results for $\sigma_{\rm el}(s)$ and $\sigma_{\rm in}(s)$ are shown in Fig.~\ref{fig:sigma_el_in}.
The model elastic cross section $\sigma_{el}(s)$ is calculated by integration of our fit to $d\sigma_{\rm el}(s,t)/dt$.
Again, we note a fair agreement with the experimental data.

\begin{figure}[tb]
\begin{center}
\includegraphics[width=0.37\textwidth]{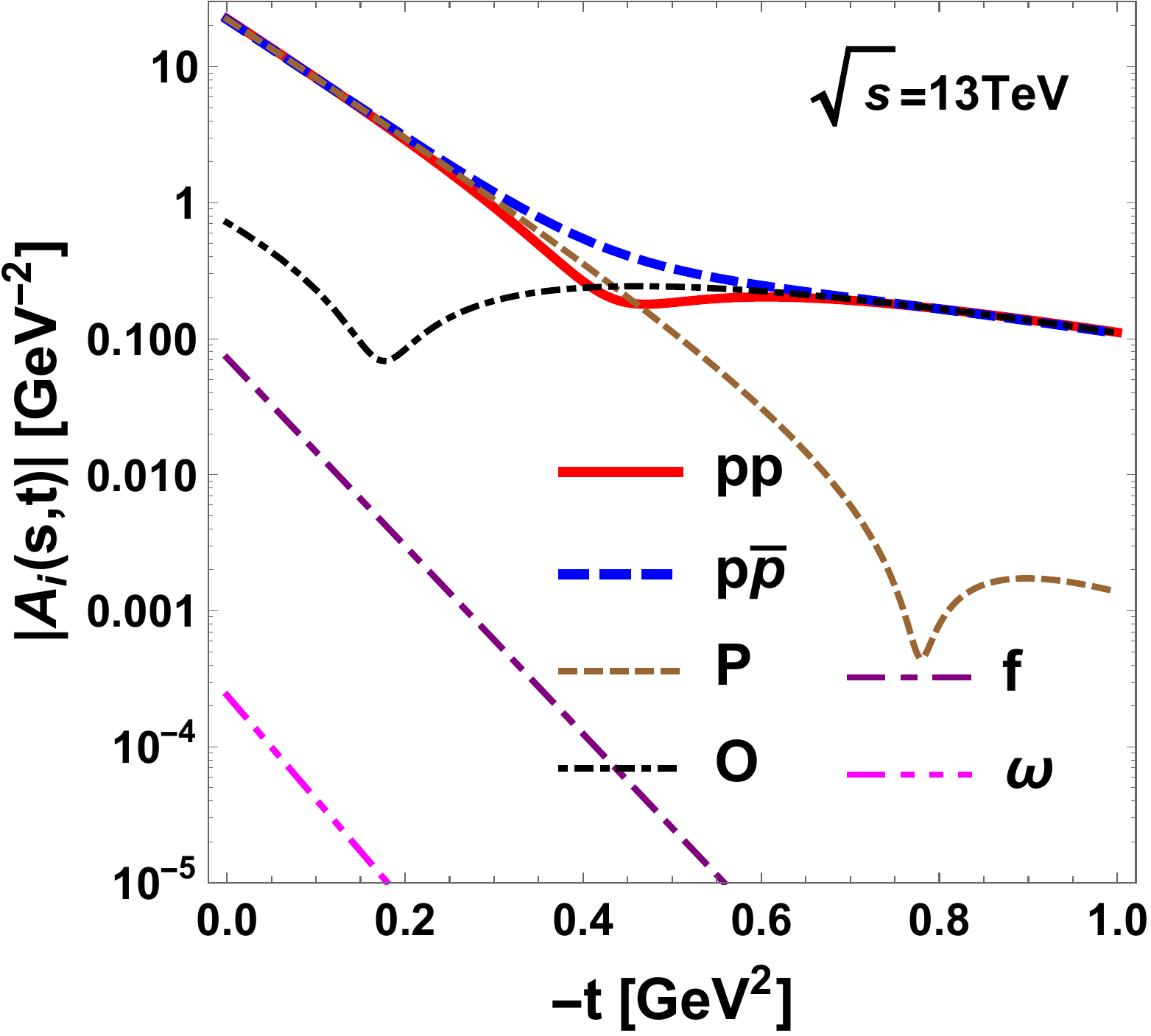}
\end{center}
\vspace{-1mm}
\caption{Absolute values of the $pp$ and $p\overline{p}$ elastic scattering amplitudes (thick lines), and absolute values of their
components (thin lines), plotted as functions of $-t$ for the TOTEM collision energy of  $\sqrt{s}=13$~TeV. At low $-t$ the Pomeron dominates, whereas at high $-t$ the odderon dominates,
and the interference of both components is manifest in the transition region around $-t=0.5~{\rm GeV}^2$. The contribution of the 
mesonic Regge trajectories is negligible at the TOTEM collision energies.
\label{fig:an_amp}}
\end{figure}

Next, we pass to a discussion of the ``anatomy'' of the model, focusing on the role of its various components. 
In Fig.~\ref{fig:an_amp} we plot the absolute values of the $pp$ and $p\overline{p}$ elastic scattering amplitudes
and their components. At low $-t$ the Pomeron contribution is dominant, and at high $-t$ the odderon takes over, as is 
evident from the relation between the slopes of their trajectories,  $\alpha'_P>\alpha'_O$, and the $b$-parameters. 
The interference of both $P$ and $O$ components shows up in the transition region around $-t=0.5~{\rm GeV}^2$, 
generating the dip. We note that the contribution of the 
mesonic Regge trajectories is negligible at the TOTEM collision energies and is essential only at low $s$.

\begin{figure}[tb]
\begin{center}
\includegraphics[width=0.37\textwidth]{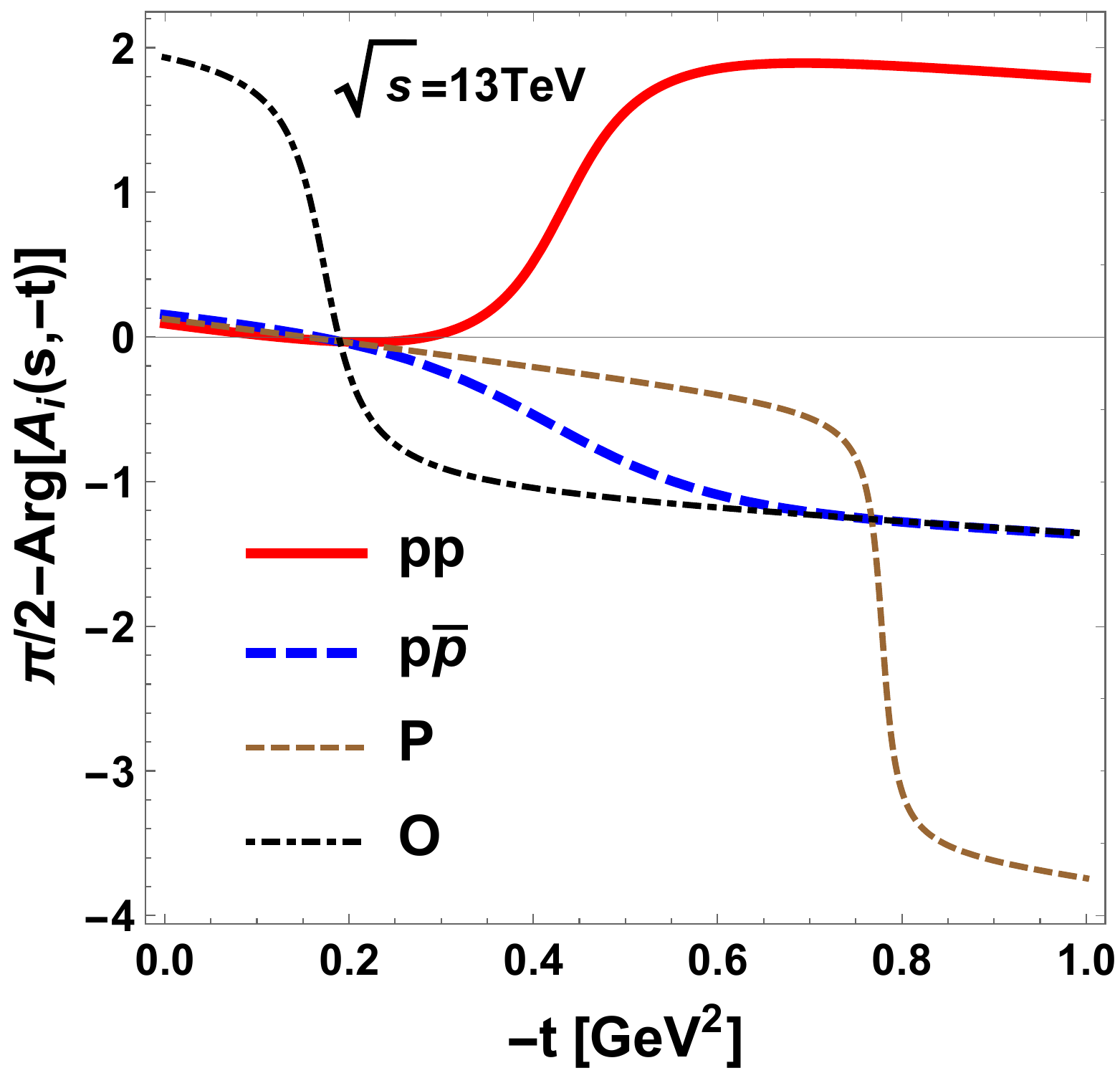}
\end{center}
\vspace{-1mm}
\caption{Same as in Fig~\ref{fig:an_amp} but for the phases, defined as \mbox{$\pi/2-{\rm Arg}[A_i(s,-t)]$}. 
At low $-t$ the phases for both $pp$ and $p\overline{p}$ are determined by the Pomeron,  
whereas at high $-t$ the phase of $p\overline{p}$ is determined by the odderon, while the 
phase of $pp$ is shifted upwards by $\pi$. \label{fig:an_ph}}
\end{figure}

In Fig.~\ref{fig:an_ph} we show the phase of the elastic amplitude, defined 
conventionally as $\pi/2-{\rm Arg}[A_i(s,-t)]$. We note that the Pomeron 
determines the phases for both $pp$ and $p\overline{p}$ at low values of  $-t$. 
At high values of $-t$ the phase of $p\overline{p}$ is determined by the odderon, and the 
phase of $pp$ is relatively shifted upwards by $\pi$, which simply reflects the relative sign between the odderon component 
of the two amplitudes.

\section{Hollowness Analysis \label{sec:hollow}}

As stated in the Introduction, in the present work we focus
on the surprising feature of the $pp$ (and $p \bar p$) scattering, the hollowness, which emerges 
at the LHC energies: the most inelastic collision become slightly peripheral, with $\sigma_{\rm in}(b,s)$ 
assuming a maximum at $b>0$, and having a minimum at $b=0$. In
what follows we complement the balanced review of Ref.~\cite{Dremin:2016ugi} and the 
discussion in~\cite{Broniowski:2017aaf}
with the results in the Regge model of Section~\ref{sec:model}. The peripheral or central
character of both the elastic and inelastic scattering was questioned
in \cite{Kundrat:1980es}, where it was shown that the shape of the
inelastic profile depends strongly on the phase of the elastic
amplitude~\cite{Kundrat:1993sv}. In a recent upgrade
\cite{Prochazka:2016wno,Prochazka:2017tby}, a preference for more peripheral 
elastic than inelastic scattering is supported, based on a careful treatment of
the Coulomb interaction and the corresponding strong phase (see,
however, the critical remarks in \cite{Petrov:2018xma}, where a different formula for the strong phase is proposed). 
The shadowing and anti-shadowing scattering scenarios have been
discussed together with the hollowness behavior in \cite{Troshin:1998ij}. In
addition, unitarization features also produce
hollowness~\cite{Cudell:2008yb}, based on the old
scheme from Ref.~\cite{Blankenbecler:1962ez}.

A recent discussion of hollowness by two of the present authors (WB
and ERA) within an inverse scattering approach, where a distinction of the
2D- and 3D-hollowness was
established~\cite{Arriola:2016bxa,RuizArriola:2016ihz,Broniowski:2017aaf,Broniowski:2017rhz},
was based on empirical parameterizations
\cite{daSilva:2008rv,Fagundes:2011jh,Fagundes:2013aja}.  These parametrization qualify as 
means of fitting the data, but actually feature no particular theory or
physical picture.
We also point out that a rather flat behavior in the inelastic profile near $b=0$ has
been observed~\cite{Selyugin:2015pha}, which may be interpreted as a
precursor of the 2D-hollowness and the occurrence of the
3D-hollowness~\cite{Arriola:2016bxa,RuizArriola:2016ihz,Broniowski:2017aaf,Broniowski:2017rhz}. Hollowness
has also been reported to emerge from a hot-spot picture of the
$pp$ collision at the LHC energies~\cite{Albacete:2016pmp}. 

In \cite{Broniowski:2017rhz} we have
shown that the existence of hollowness depends strongly on the $t$-dependence of the
$\rho$ parameter~\footnote{The criticism that has been raised in
Ref.~\cite{Petrov:2018wlv} is based on an incorrect perception of the
approximations involved and does not address the arbitrariness of the
$t$-dependence of the $\rho(s,t)$ parameter, which is crucial for hollowness.}.  Once we recognize that
hollowness cannot be deduced from present data alone, in this paper we
take a different point of view, where we want to decide on the hollowness
within a given theoretical framework.  

\begin{figure}[tb]
\begin{center}
\includegraphics[width=0.37\textwidth]{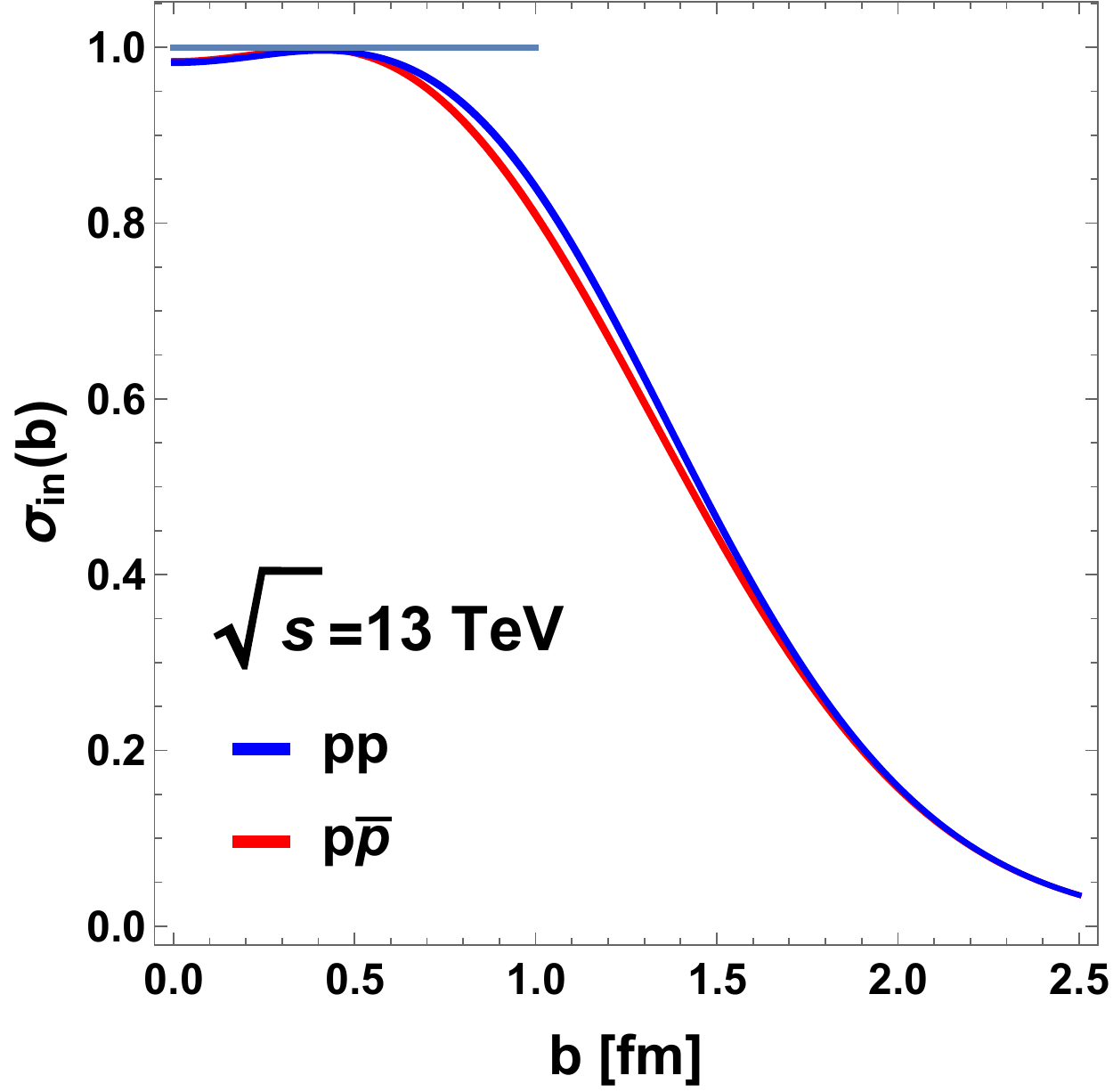}
\end{center}
\vspace{-1mm}
\caption{The inelastic profiles $\sigma(b,s)$ for $pp$ and $p\overline{p}$ collisions, plotted as 
functions of the impact parameter $b$ for the TOTEM collision energy of  $\sqrt{s}=13$~TeV. We note the presence of the hollowness effect, 
somewhat stronger for  $p\overline{p}$ than for  $pp$.  The uncertainty bands, following from the error propagation from
our fit parameters, are within the thickness of the lines. \label{fig:hollow}}
\end{figure}

\begin{figure}[tb]
\begin{center}
\includegraphics[width=0.37\textwidth]{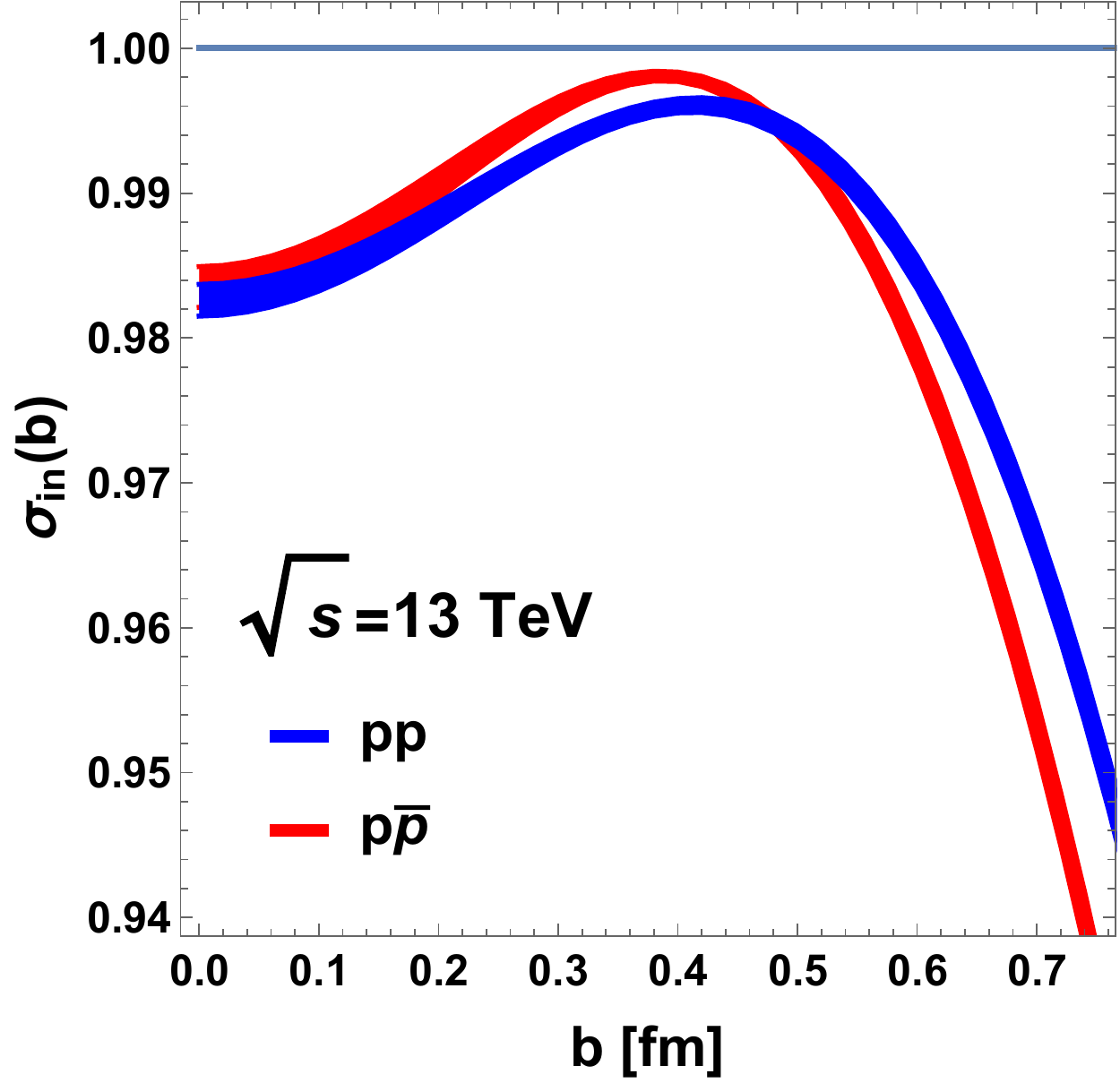}
\end{center}
\vspace{-1mm}
\caption{Close-up of Fig.~\ref{fig:hollow}, with visible error bands.  \label{fig:hollow_c}}
\end{figure}

In the following, we apply the formulas of Subsection~\ref{sec:imp} to
the model of Section~\ref{sec:model}.  We first look at the
inelasticity profile $\sigma_{\rm in}(b,s)$, shown $\sqrt{s}=13$~TeV
in Figs.~\ref{fig:hollow} and \ref{fig:hollow_c}.  We clearly note the
feature of hollowness, i.e., a (shallow) minimum in the center. We
note that the phenomenon occurs for both $pp$ and $p \bar p$
collisions, and is slightly stronger for the latter case. We also
display the uncertainty bands corresponding to the error propagation from
our fit parameters using the conventional error
matrix~\footnote{Namely, for a function $F(p_1, \dots , p_N)$ of the
fitting parameters $p_i$ with correlation matrix ${\cal C}_{ij}$ we
take $(\Delta F)^2 = \sum_{ij} {\cal C}_{ij} \partial_i F \partial_j F 
$ }. As we can clearly see the hollowness is a robust feature as long
as the statistical uncertainties are concerned. 

\begin{figure}[tb]
\begin{center}
\includegraphics[width=0.37\textwidth]{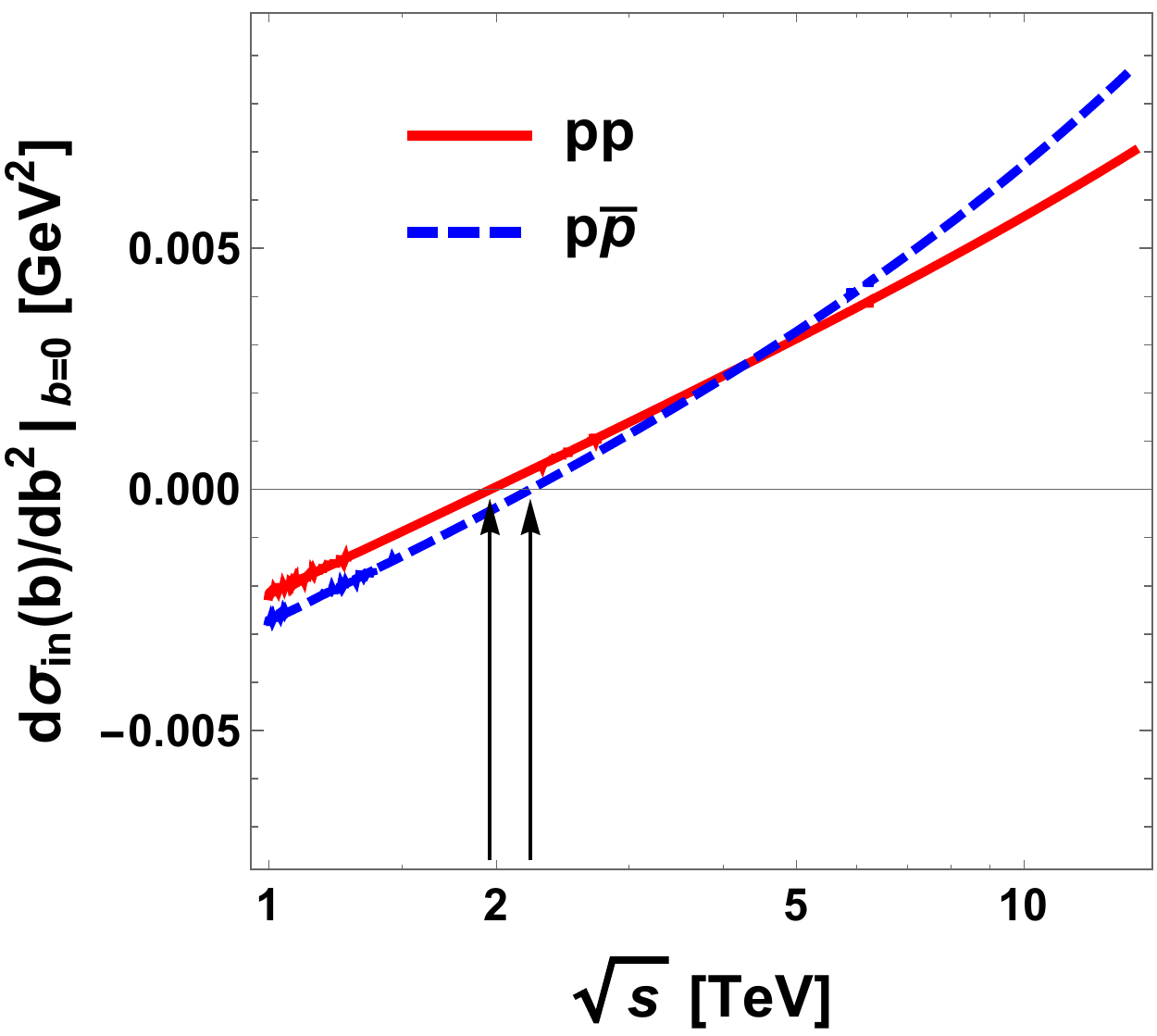}
\end{center}
\vspace{-1mm}
\caption{The criterion for hollowness for $pp$ and $p\overline{p}$, Eq.~(\ref{eq:crit}), plotted as a function of the collision energy. 
Positive values of the curves mean hollowness, with its onset indicated with arrows. 
\label{fig:cond}}
\end{figure}

From Fig.~\ref{fig:cond} we see that the onset of hollowness occurs at
similar collision energies for $pp$ and for $p\overline{p}$, namely around
3~TeV.  However, at higher $\sqrt{s}$ the hollowness becomes somewhat stronger
in $p\overline{p}$ compared to $pp$, as the curvature at the origin is
larger for the former case (the dashed curve goes above the solid
curve in Fig.~\ref{fig:cond}). This is also manifest in the behavior
displayed in Fig.~\ref{fig:hollow_c}.

\begin{figure}[tb]
\begin{center}
\includegraphics[width=0.37\textwidth]{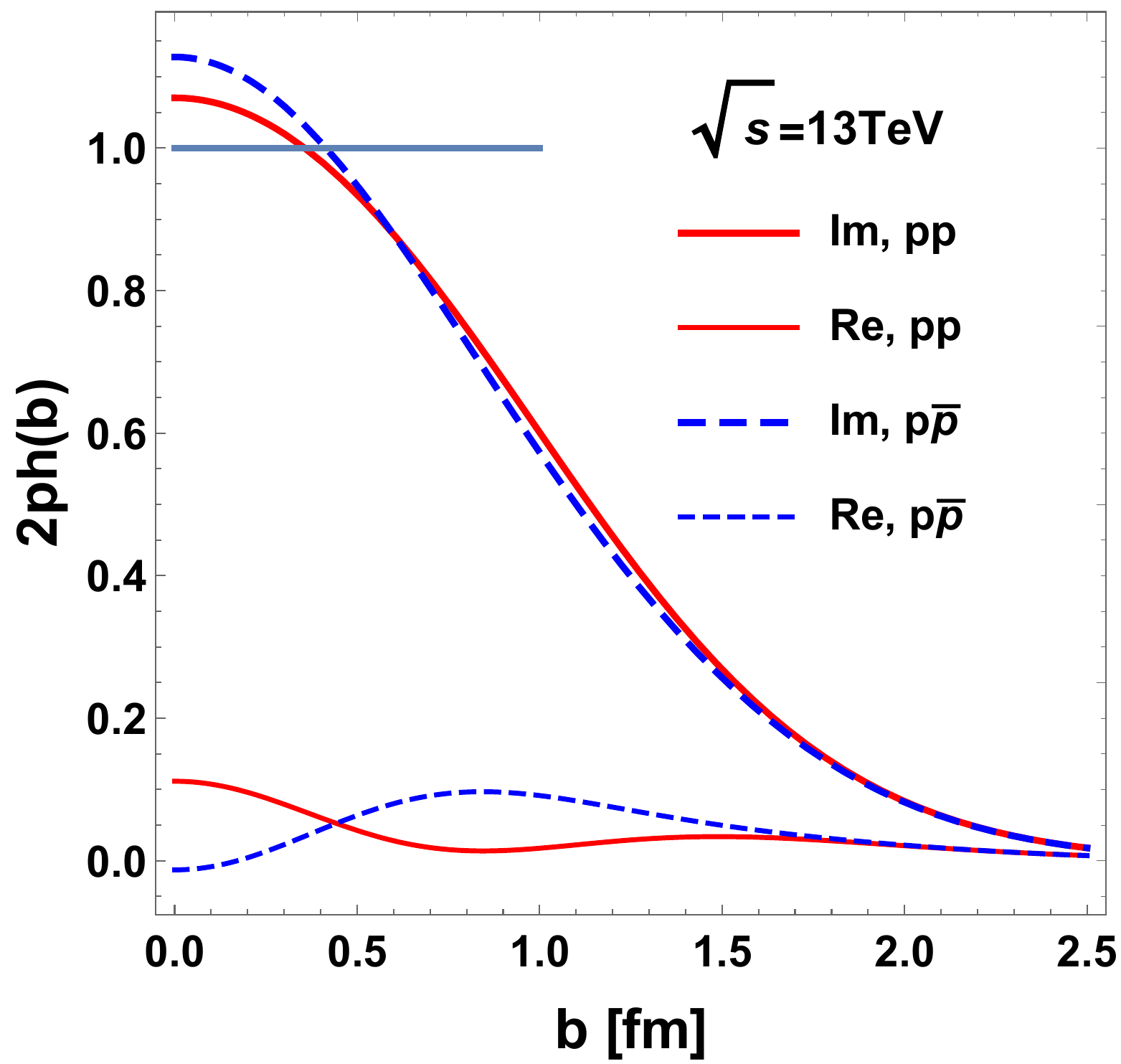}
\end{center}
\vspace{-1mm}
\caption{The real and imaginary parts of the amplitudes $2p\, h(b,s)$ for $pp$ and $p\overline{p}$, plotted as 
functions of the impact parameter $b$ for the TOTEM collision energy of  $\sqrt{s}=13$~TeV. We note that 
near the origin the imaginary parts go above 1. \label{fig:h}}
\end{figure}

In the model with strictly linear Regge trajectories the amplitude is
a combination of Gaussians in $q^2$, hence the expression for
$2p\,h(b)$ is analytic, involving a combination of Gaussians in
$b$ (with complex parameters).  In that case one can write down the
criterion of Eq.~(\ref{eq:crit}) for hollowness in terms of a relation
of the model parameters and $s$. However, the final formula is long and not
very illuminating.  A simpler
result follows with the condition
\begin{eqnarray}
2p\,{\rm Im} \,h(s,b=0) > 1, \label{eq:cond1}
\end{eqnarray}
which becomes equivalent to Eq.~(\ref{eq:crit}) in the absence of the real part in the amplitude~\cite{RuizArriola:2016ihz,Broniowski:2017aaf}. 
The behavior of the real and imaginary parts of the elastic amplitude in the $b$-representation, $2p\,h(b,s)$, 
plotted in Fig.~\ref{fig:h}. We note that 
near $b=0$ the imaginary parts go above 1, whereas the real parts are small.

\begin{figure}[tb]
\begin{center}
\includegraphics[width=0.37\textwidth]{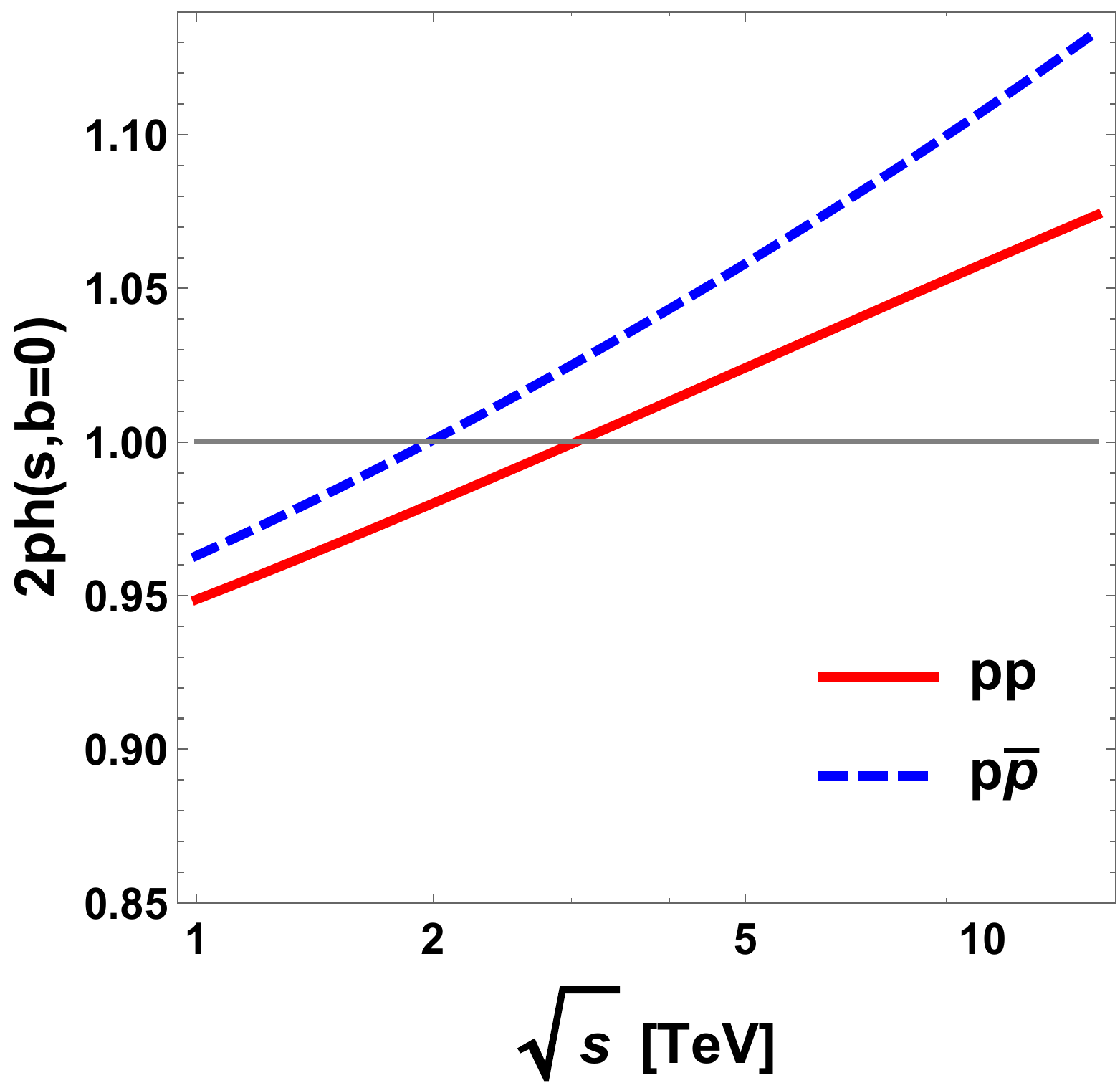}
\end{center}
\vspace{-1mm} 
\caption{Value of $2p\,{\rm Im}h$ at the origin, plotted as a function of $\sqrt{s}$. 
The curves cross the value 1 near the hollowness transition at $\sqrt{s} \sim 3$~TeV. \label{fig:hor}}
\end{figure}

In the model with strictly linear trajectories we find numerically for the dominant component
\begin{eqnarray}
2p\,{\rm Im} \,h_P(b=0,s)&=&\frac{a_P \cos \left(\frac{\pi  \delta _P}{2}\right) 
\left(e^{b_P \delta _P}-\epsilon _P\right) \left(\frac{s}{s_{{P0}}}\right){}^{\delta _P}}{2 b_P s_{{P0}} \alpha _{{P1}}} \nonumber \\
                    &\simeq& 0.64 (s/{\rm GeV}^2)^{0.028},
\nonumber  \\ 
2p\,{\rm Im}\, h_O(b=0,s)&=&\frac{a_O \sin \left(\frac{\pi  \delta _O}{2}\right) 
\left(e^{\epsilon _O-b_O \delta _O}\right) \left(\frac{s}{s_{{O0}}}\right){}^{\delta _O}}{2 b_O \alpha _{{O1}} s_{{O1}}} \nonumber \\
                    &\simeq& 0.88 (s/{\rm GeV}^2)^{0.27}. 
\end{eqnarray}
The growth with $s$ leads to inevitable crossing of the value 1 at $b=0$, as seen in Fig.~\ref{fig:hor}.

\begin{figure}[tb]
\begin{center}
\includegraphics[width=0.37\textwidth]{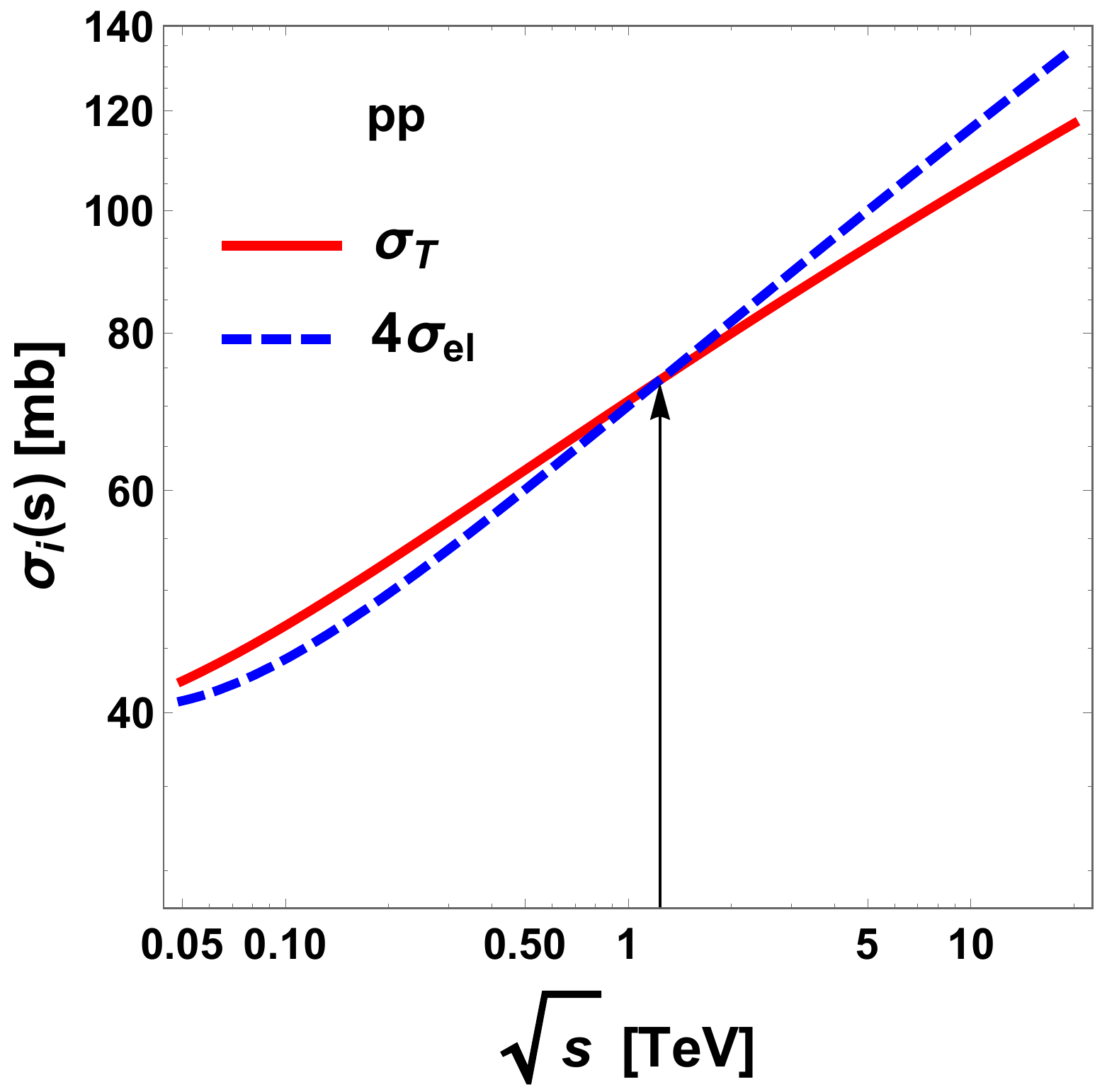}
\end{center}
\vspace{-1mm}
\caption{
Comparison of the total and 4 times the elastic $pp$ cross sections, plotted as a function of $\sqrt{s}$.
The crossing occurs near the transition to hollowness at $\sqrt{s} \sim 3$~TeV.
\label{fig:spp4}}
\end{figure}

\begin{figure}[tb]
\begin{center}
\includegraphics[width=0.37\textwidth]{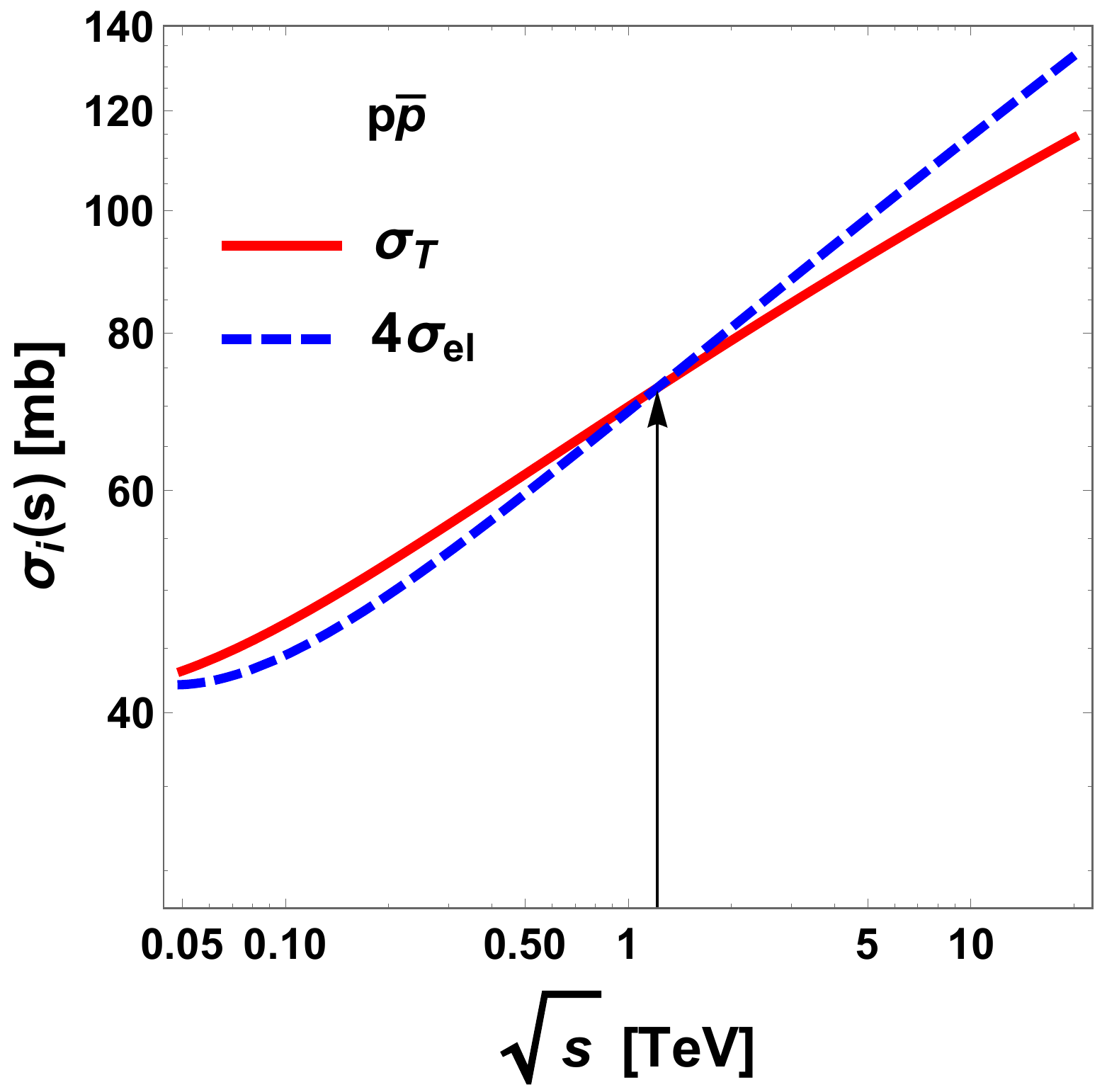}
\end{center}
\vspace{-1mm} 
\caption{Same as Fig.~\ref{fig:spp4}, but for $p\bar p$ scattering
\label{fig:sppb4}}
\end{figure}

Finally, we examine condition~(\ref{eq:to}) in our model. The result for the total and 4 times the elastic $pp$ cross sections
is displayed in Figs.~\ref{fig:spp4} and \ref{fig:sppb4} for the $pp$ and $p\bar p$ scattering, respectively. We notice the expected crossing near the 
hollowness transition, near $\sqrt{s} \sim 3$~TeV.

We have thus demonstrated a firm occurrence of hollowness in the Regge model of Section~\ref{sec:model} above
$\sqrt{s} \sim 3$~TeV.  We have also illustrated the criteria for its appearance. 
As shown in~\cite{RuizArriola:2016ihz}, 
the very existence of the hollowness phenomenon is quantum-mechanical in nature, as 
it invalidates folding constructions of the inelasticity profile, as used e.g. in 
Refs.~\cite{Chou:1968bc,Chou:1968bg,Cheng:1987ga,Bourrely:1978da,Block:2006hy,Block:2015sea}, where hollowness is prevented from the outset.

After this paper was submitted, a work analyzing the overlap
function for the pp TOTEM data at the fixed CM energy of 13
TeV~\cite{Alkin:2018vhh} was released. It uses the finite binning method suggested
in~\cite{Amaldi:1979kd}, including also error estimates. The results of~\cite{Alkin:2018vhh}
are compatible with ours, with a clear development of the hollow. We
recall that ours is a multi-energy analysis which includes both $pp$ and
$p\bar p$ within a Regge setup. The persistent occurrence of
hollowness at the LHC can be traced to quantum mechanical
interference effects which defy purely geometric
models~\cite{RuizArriola:2016ihz}. As we have already mentioned, our
motivation to use the Regge approach is a realization of analytic
properties, such as a fixed-$t$ dispersion relation and crossing, which
allows to fix the phase of the amplitude whose absolute value is
determined from LHC data. The question whether it is possible or not
to redesign a model where analyticity and geometric features are
implemented and at the same time data are described, is left for future
research.

\section{Conclusions \label{sec:concl}}

Over the years, there have been two basic, presumably complementary
approaches to the high-energy hadron scattering. The geometric and
Regge models rest on different assumptions but they have been
implicitly assumed to be dual to each other~\cite{Barone:2002cv} in
the sense that they emphasize the $t$ or $s$ dependence of the
scattering amplitude. Based on the results shown in this paper, we
argue that this may not necessarily be so, as a simple Regge model
fitting globally the $pp$ and $p \bar p$ data in a wide energy range
displays hollowness at the LHC energies, an effect incompatible with
the folding feature characterizing the geometric
models~\cite{RuizArriola:2016ihz}.  At the same time hollowness is not
a priori precluded by the Regge theory nor by the fixed-$t$ dispersion
relations, and our analysis exemplifies this possibility.

Admittedly, the emergence of hollowness, in $pp$ and $p \bar p$
collisions is a remarkable and unexpected feature, unveiling a so far
puzzling property which can only stem from a quantum mechanical effect
encoded in the amplitude. However, this property cannot be deduced
solely from the elastic scattering data, but needs some theoretical
assumptions enabling to constrain the phase. Our Regge model includes
both the Pomeron and the odderon on equal footing, which properly
describes the energy- and momentum-transfer dependence in the range
from the old ISR to the newest LHC data, and exhibits the phenomenon
of hollowness. We have shown that within this framework the onset of
hollowness rests mainly on the double-pole Pomeron component. Within
the present model we predict the transition region to take place at
$\sqrt{s}\sim 3$~TeV, and the feature to hold at all higher energies.
Further work is needed to unveil the precise microscopic
mechanism behind this emergent and intriguing phenomenon.

\begin{acknowledgments}
This research was supported by the Polish National
Science Center, Grant 2015/19/B/ST2/00937 (WB), by the Spanish Mineco Grants FIS2014-59386-P and FIS2017-85053-C2-1-P (ERA), by the
Junta de Andaluc\'{\i}a, Grant FQM225-05 (ERA), and by the Ukrainian National Academy's grant ``Structure and dynamics of  statistic and quantum-mechanical systems'' (LJ). 
\end{acknowledgments}

\bibliographystyle{apsrev4-1}

\bibliography{NN-high-energy,PomOdd}

\begin{thebibliography}{74}%
\makeatletter
\providecommand \@ifxundefined [1]{%
 \@ifx{#1\undefined}
}%
\providecommand \@ifnum [1]{%
 \ifnum #1\expandafter \@firstoftwo
 \else \expandafter \@secondoftwo
 \fi
}%
\providecommand \@ifx [1]{%
 \ifx #1\expandafter \@firstoftwo
 \else \expandafter \@secondoftwo
 \fi
}%
\providecommand \natexlab [1]{#1}%
\providecommand \enquote  [1]{``#1''}%
\providecommand \bibnamefont  [1]{#1}%
\providecommand \bibfnamefont [1]{#1}%
\providecommand \citenamefont [1]{#1}%
\providecommand \href@noop [0]{\@secondoftwo}%
\providecommand \href [0]{\begingroup \@sanitize@url \@href}%
\providecommand \@href[1]{\@@startlink{#1}\@@href}%
\providecommand \@@href[1]{\endgroup#1\@@endlink}%
\providecommand \@sanitize@url [0]{\catcode `\\12\catcode `\$12\catcode
  `\&12\catcode `\#12\catcode `\^12\catcode `\_12\catcode `\%12\relax}%
\providecommand \@@startlink[1]{}%
\providecommand \@@endlink[0]{}%
\providecommand \url  [0]{\begingroup\@sanitize@url \@url }%
\providecommand \@url [1]{\endgroup\@href {#1}{\urlprefix }}%
\providecommand \urlprefix  [0]{URL }%
\providecommand \Eprint [0]{\href }%
\providecommand \doibase [0]{http://dx.doi.org/}%
\providecommand \selectlanguage [0]{\@gobble}%
\providecommand \bibinfo  [0]{\@secondoftwo}%
\providecommand \bibfield  [0]{\@secondoftwo}%
\providecommand \translation [1]{[#1]}%
\providecommand \BibitemOpen [0]{}%
\providecommand \bibitemStop [0]{}%
\providecommand \bibitemNoStop [0]{.\EOS\space}%
\providecommand \EOS [0]{\spacefactor3000\relax}%
\providecommand \BibitemShut  [1]{\csname bibitem#1\endcsname}%
\let\auto@bib@innerbib\@empty
\bibitem [{\citenamefont {Barone}\ and\ \citenamefont
  {Predazzi}(2002)}]{Barone:2002cv}%
  \BibitemOpen
  \bibfield  {author} {\bibinfo {author} {\bibfnamefont {V.}~\bibnamefont
  {Barone}}\ and\ \bibinfo {author} {\bibfnamefont {E.}~\bibnamefont
  {Predazzi}},\ }\href@noop {} {\emph {\bibinfo {title} {{High-Energy Particle
  Diffraction}}}}\ (\bibinfo  {publisher} {Springer-Verlag},\ \bibinfo
  {address} {Berlin Heidelberg},\ \bibinfo {year} {2002})\BibitemShut {NoStop}%
\bibitem [{\citenamefont {Dremin}(2013)}]{Dremin:2012ke}%
  \BibitemOpen
  \bibfield  {author} {\bibinfo {author} {\bibfnamefont {I.~M.}\ \bibnamefont
  {Dremin}},\ }\href {\doibase 10.3367/UFNe.0183.201301a.0003} {\bibfield
  {journal} {\bibinfo  {journal} {Phys. Usp.}\ }\textbf {\bibinfo {volume}
  {56}},\ \bibinfo {pages} {3} (\bibinfo {year} {2013})},\ \bibinfo {note}
  {[Usp. Fiz. Nauk 183, 3 (2013)]},\ \Eprint {http://arxiv.org/abs/1206.5474}
  {arXiv:1206.5474 [hep-ph]} \BibitemShut {NoStop}%
\bibitem [{\citenamefont {Pancheri}\ and\ \citenamefont
  {Srivastava}(2017)}]{Pancheri:2016yel}%
  \BibitemOpen
  \bibfield  {author} {\bibinfo {author} {\bibfnamefont {G.}~\bibnamefont
  {Pancheri}}\ and\ \bibinfo {author} {\bibfnamefont {Y.~N.}\ \bibnamefont
  {Srivastava}},\ }\href {\doibase 10.1140/epjc/s10052-016-4585-8} {\bibfield
  {journal} {\bibinfo  {journal} {Eur. Phys. J.}\ }\textbf {\bibinfo {volume}
  {C77}},\ \bibinfo {pages} {150} (\bibinfo {year} {2017})},\ \Eprint
  {http://arxiv.org/abs/1610.10038} {arXiv:1610.10038 [hep-ph]} \BibitemShut
  {NoStop}%
\bibitem [{\citenamefont {Van~Hove}(1963)}]{van1963phenomenological}%
  \BibitemOpen
  \bibfield  {author} {\bibinfo {author} {\bibfnamefont {L.}~\bibnamefont
  {Van~Hove}},\ }\href@noop {} {\bibfield  {journal} {\bibinfo  {journal} {Il
  Nuovo Cimento (1955-1965)}\ }\textbf {\bibinfo {volume} {28}},\ \bibinfo
  {pages} {798} (\bibinfo {year} {1963})}\BibitemShut {NoStop}%
\bibitem [{\citenamefont {Van~Hove}(1964)}]{VanHove:1964rp}%
  \BibitemOpen
  \bibfield  {author} {\bibinfo {author} {\bibfnamefont {L.}~\bibnamefont
  {Van~Hove}},\ }\href {\doibase 10.1103/RevModPhys.36.655} {\bibfield
  {journal} {\bibinfo  {journal} {Rev. Mod. Phys.}\ }\textbf {\bibinfo {volume}
  {36}},\ \bibinfo {pages} {655} (\bibinfo {year} {1964})}\BibitemShut
  {NoStop}%
\bibitem [{\citenamefont {Amaldi}\ and\ \citenamefont
  {Schubert}(1980)}]{Amaldi:1979kd}%
  \BibitemOpen
  \bibfield  {author} {\bibinfo {author} {\bibfnamefont {U.}~\bibnamefont
  {Amaldi}}\ and\ \bibinfo {author} {\bibfnamefont {K.~R.}\ \bibnamefont
  {Schubert}},\ }\href {\doibase 10.1016/0550-3213(80)90229-1} {\bibfield
  {journal} {\bibinfo  {journal} {Nucl.Phys.}\ }\textbf {\bibinfo {volume}
  {B166}},\ \bibinfo {pages} {301} (\bibinfo {year} {1980})}\BibitemShut
  {NoStop}%
\bibitem [{\citenamefont {Chou}\ and\ \citenamefont
  {Yang}(1968{\natexlab{a}})}]{Chou:1968bc}%
  \BibitemOpen
  \bibfield  {author} {\bibinfo {author} {\bibfnamefont {T.~T.}\ \bibnamefont
  {Chou}}\ and\ \bibinfo {author} {\bibfnamefont {C.-N.}\ \bibnamefont
  {Yang}},\ }\href {\doibase 10.1103/PhysRev.170.1591} {\bibfield  {journal}
  {\bibinfo  {journal} {Phys. Rev.}\ }\textbf {\bibinfo {volume} {170}},\
  \bibinfo {pages} {1591} (\bibinfo {year} {1968}{\natexlab{a}})}\BibitemShut
  {NoStop}%
\bibitem [{\citenamefont {Chou}\ and\ \citenamefont
  {Yang}(1968{\natexlab{b}})}]{Chou:1968bg}%
  \BibitemOpen
  \bibfield  {author} {\bibinfo {author} {\bibfnamefont {T.~T.}\ \bibnamefont
  {Chou}}\ and\ \bibinfo {author} {\bibfnamefont {C.-N.}\ \bibnamefont
  {Yang}},\ }\href {\doibase 10.1103/PhysRev.175.1832} {\bibfield  {journal}
  {\bibinfo  {journal} {Phys. Rev.}\ }\textbf {\bibinfo {volume} {175}},\
  \bibinfo {pages} {1832} (\bibinfo {year} {1968}{\natexlab{b}})}\BibitemShut
  {NoStop}%
\bibitem [{\citenamefont {Cheng}\ and\ \citenamefont
  {Wu}(1987)}]{Cheng:1987ga}%
  \BibitemOpen
  \bibfield  {author} {\bibinfo {author} {\bibfnamefont {H.}~\bibnamefont
  {Cheng}}\ and\ \bibinfo {author} {\bibfnamefont {T.~T.}\ \bibnamefont {Wu}},\
  }\href@noop {} {\emph {\bibinfo {title} {Expanding protons: Scattering at
  high energies}}}\ (\bibinfo  {publisher} {{MIT Press}},\ \bibinfo {year}
  {1987})\BibitemShut {NoStop}%
\bibitem [{\citenamefont {Bourrely}\ \emph {et~al.}(1979)\citenamefont
  {Bourrely}, \citenamefont {Soffer},\ and\ \citenamefont
  {Wu}}]{Bourrely:1978da}%
  \BibitemOpen
  \bibfield  {author} {\bibinfo {author} {\bibfnamefont {C.}~\bibnamefont
  {Bourrely}}, \bibinfo {author} {\bibfnamefont {J.}~\bibnamefont {Soffer}}, \
  and\ \bibinfo {author} {\bibfnamefont {T.~T.}\ \bibnamefont {Wu}},\ }\href
  {\doibase 10.1103/PhysRevD.19.3249} {\bibfield  {journal} {\bibinfo
  {journal} {Phys. Rev.}\ }\textbf {\bibinfo {volume} {D19}},\ \bibinfo {pages}
  {3249} (\bibinfo {year} {1979})}\BibitemShut {NoStop}%
\bibitem [{\citenamefont {Block}(2006)}]{Block:2006hy}%
  \BibitemOpen
  \bibfield  {author} {\bibinfo {author} {\bibfnamefont {M.~M.}\ \bibnamefont
  {Block}},\ }\href {\doibase 10.1016/j.physrep.2006.06.003} {\bibfield
  {journal} {\bibinfo  {journal} {Phys. Rept.}\ }\textbf {\bibinfo {volume}
  {436}},\ \bibinfo {pages} {71} (\bibinfo {year} {2006})},\ \Eprint
  {http://arxiv.org/abs/hep-ph/0606215} {arXiv:hep-ph/0606215 [hep-ph]}
  \BibitemShut {NoStop}%
\bibitem [{\citenamefont {Block}\ \emph {et~al.}(2015)\citenamefont {Block},
  \citenamefont {Durand}, \citenamefont {Ha},\ and\ \citenamefont
  {Halzen}}]{Block:2015sea}%
  \BibitemOpen
  \bibfield  {author} {\bibinfo {author} {\bibfnamefont {M.~M.}\ \bibnamefont
  {Block}}, \bibinfo {author} {\bibfnamefont {L.}~\bibnamefont {Durand}},
  \bibinfo {author} {\bibfnamefont {P.}~\bibnamefont {Ha}}, \ and\ \bibinfo
  {author} {\bibfnamefont {F.}~\bibnamefont {Halzen}},\ }\href {\doibase
  10.1103/PhysRevD.92.014030} {\bibfield  {journal} {\bibinfo  {journal} {Phys.
  Rev.}\ }\textbf {\bibinfo {volume} {D92}},\ \bibinfo {pages} {014030}
  (\bibinfo {year} {2015})},\ \Eprint {http://arxiv.org/abs/1505.04842}
  {arXiv:1505.04842 [hep-ph]} \BibitemShut {NoStop}%
\bibitem [{\citenamefont {Alkin}\ \emph {et~al.}(2014)\citenamefont {Alkin},
  \citenamefont {Martynov}, \citenamefont {Kovalenko},\ and\ \citenamefont
  {Troshin}}]{Alkin:2014rfa}%
  \BibitemOpen
  \bibfield  {author} {\bibinfo {author} {\bibfnamefont {A.}~\bibnamefont
  {Alkin}}, \bibinfo {author} {\bibfnamefont {E.}~\bibnamefont {Martynov}},
  \bibinfo {author} {\bibfnamefont {O.}~\bibnamefont {Kovalenko}}, \ and\
  \bibinfo {author} {\bibfnamefont {S.~M.}\ \bibnamefont {Troshin}},\ }\href
  {\doibase 10.1103/PhysRevD.89.091501} {\bibfield  {journal} {\bibinfo
  {journal} {Phys. Rev.}\ }\textbf {\bibinfo {volume} {D89}},\ \bibinfo {pages}
  {091501} (\bibinfo {year} {2014})},\ \Eprint {http://arxiv.org/abs/1403.8036}
  {arXiv:1403.8036 [hep-ph]} \BibitemShut {NoStop}%
\bibitem [{\citenamefont {Dremin}(2015{\natexlab{a}})}]{Dremin:2014eva}%
  \BibitemOpen
  \bibfield  {author} {\bibinfo {author} {\bibfnamefont {I.~M.}\ \bibnamefont
  {Dremin}},\ }\href {\doibase 10.3103/S1068335615010066} {\bibfield  {journal}
  {\bibinfo  {journal} {Bull. Lebedev Phys. Inst.}\ }\textbf {\bibinfo {volume}
  {42}},\ \bibinfo {pages} {21} (\bibinfo {year} {2015}{\natexlab{a}})},\
  \bibinfo {note} {[Kratk. Soobshch. Fiz. 42, no. 1, 8 (2015)]},\ \Eprint
  {http://arxiv.org/abs/1404.4142} {arXiv:1404.4142 [hep-ph]} \BibitemShut
  {NoStop}%
\bibitem [{\citenamefont {Dremin}(2015{\natexlab{b}})}]{Dremin:2014spa}%
  \BibitemOpen
  \bibfield  {author} {\bibinfo {author} {\bibfnamefont {I.~M.}\ \bibnamefont
  {Dremin}},\ }\href {\doibase 10.3367/UFNe.0185.201501d.0065} {\bibfield
  {journal} {\bibinfo  {journal} {Phys. Usp.}\ }\textbf {\bibinfo {volume}
  {58}},\ \bibinfo {pages} {61} (\bibinfo {year} {2015}{\natexlab{b}})},\
  \Eprint {http://arxiv.org/abs/1406.2153} {arXiv:1406.2153 [hep-ph]}
  \BibitemShut {NoStop}%
\bibitem [{\citenamefont {Ruiz~Arriola}\ and\ \citenamefont
  {Broniowski}(2016)}]{Arriola:2016bxa}%
  \BibitemOpen
  \bibfield  {author} {\bibinfo {author} {\bibfnamefont {E.}~\bibnamefont
  {Ruiz~Arriola}}\ and\ \bibinfo {author} {\bibfnamefont {W.}~\bibnamefont
  {Broniowski}},\ }\bibfield  {booktitle} {\emph {\bibinfo {booktitle}
  {{Proceedings, Theory and Experiment for Hadrons on the Light-Front (Light
  Cone 2015): Frascati , Italy, September 21-25, 2015}}},\ }\href {\doibase
  10.1007/s00601-016-1095-z} {\bibfield  {journal} {\bibinfo  {journal} {Few
  Body Syst.}\ }\textbf {\bibinfo {volume} {57}},\ \bibinfo {pages} {485}
  (\bibinfo {year} {2016})},\ \Eprint {http://arxiv.org/abs/1602.00288}
  {arXiv:1602.00288 [hep-ph]} \BibitemShut {NoStop}%
\bibitem [{\citenamefont {Ruiz~Arriola}\ and\ \citenamefont
  {Broniowski}(2017)}]{RuizArriola:2016ihz}%
  \BibitemOpen
  \bibfield  {author} {\bibinfo {author} {\bibfnamefont {E.}~\bibnamefont
  {Ruiz~Arriola}}\ and\ \bibinfo {author} {\bibfnamefont {W.}~\bibnamefont
  {Broniowski}},\ }\href {\doibase 10.1103/PhysRevD.95.074030} {\bibfield
  {journal} {\bibinfo  {journal} {Phys. Rev.}\ }\textbf {\bibinfo {volume}
  {D95}},\ \bibinfo {pages} {074030} (\bibinfo {year} {2017})},\ \Eprint
  {http://arxiv.org/abs/1609.05597} {arXiv:1609.05597 [nucl-th]} \BibitemShut
  {NoStop}%
\bibitem [{\citenamefont {Albacete}\ and\ \citenamefont
  {Soto-Ontoso}(2017)}]{Albacete:2016pmp}%
  \BibitemOpen
  \bibfield  {author} {\bibinfo {author} {\bibfnamefont {J.~L.}\ \bibnamefont
  {Albacete}}\ and\ \bibinfo {author} {\bibfnamefont {A.}~\bibnamefont
  {Soto-Ontoso}},\ }\href {\doibase 10.1016/j.physletb.2017.04.055} {\bibfield
  {journal} {\bibinfo  {journal} {Phys. Lett.}\ }\textbf {\bibinfo {volume}
  {B770}},\ \bibinfo {pages} {149} (\bibinfo {year} {2017})},\ \Eprint
  {http://arxiv.org/abs/1605.09176} {arXiv:1605.09176 [hep-ph]} \BibitemShut
  {NoStop}%
\bibitem [{\citenamefont {Dremin}(2017{\natexlab{a}})}]{Dremin:2016ugi}%
  \BibitemOpen
  \bibfield  {author} {\bibinfo {author} {\bibfnamefont {I.~M.}\ \bibnamefont
  {Dremin}},\ }\href {\doibase 10.3367/UFNe.2016.11.037977} {\bibfield
  {journal} {\bibinfo  {journal} {Phys. Usp.}\ }\textbf {\bibinfo {volume}
  {60}},\ \bibinfo {pages} {333} (\bibinfo {year} {2017}{\natexlab{a}})},\
  \Eprint {http://arxiv.org/abs/1610.07937} {arXiv:1610.07937 [hep-ph]}
  \BibitemShut {NoStop}%
\bibitem [{\citenamefont {Broniowski}\ and\ \citenamefont
  {Ruiz~Arriola}(2017{\natexlab{a}})}]{Broniowski:2017aaf}%
  \BibitemOpen
  \bibfield  {author} {\bibinfo {author} {\bibfnamefont {W.}~\bibnamefont
  {Broniowski}}\ and\ \bibinfo {author} {\bibfnamefont {E.}~\bibnamefont
  {Ruiz~Arriola}},\ }\bibfield  {booktitle} {\emph {\bibinfo {booktitle}
  {{Proceedings, 23rd Cracow Epiphany Conference on Particle Theory Meets the
  First Data from LHC Run 2: Cracow, Poland, January 9-12, 2017}}},\ }\href
  {\doibase 10.5506/APhysPolB.48.927} {\bibfield  {journal} {\bibinfo
  {journal} {Acta Phys. Polon.}\ }\textbf {\bibinfo {volume} {B48}},\ \bibinfo
  {pages} {927} (\bibinfo {year} {2017}{\natexlab{a}})},\ \Eprint
  {http://arxiv.org/abs/1704.03271} {arXiv:1704.03271 [hep-ph]} \BibitemShut
  {NoStop}%
\bibitem [{\citenamefont {Dremin}(2017{\natexlab{b}})}]{Dremin:2017ylm}%
  \BibitemOpen
  \bibfield  {author} {\bibinfo {author} {\bibfnamefont {I.~M.}\ \bibnamefont
  {Dremin}},\ }\href@noop {} {\  (\bibinfo {year} {2017}{\natexlab{b}})},\
  \Eprint {http://arxiv.org/abs/1702.06304} {arXiv:1702.06304 [hep-ph]}
  \BibitemShut {NoStop}%
\bibitem [{\citenamefont {Anisovich}\ \emph {et~al.}(2014)\citenamefont
  {Anisovich}, \citenamefont {Nikonov},\ and\ \citenamefont
  {Nyiri}}]{Anisovich:2014wha}%
  \BibitemOpen
  \bibfield  {author} {\bibinfo {author} {\bibfnamefont {V.~V.}\ \bibnamefont
  {Anisovich}}, \bibinfo {author} {\bibfnamefont {V.~A.}\ \bibnamefont
  {Nikonov}}, \ and\ \bibinfo {author} {\bibfnamefont {J.}~\bibnamefont
  {Nyiri}},\ }\href {\doibase 10.1103/PhysRevD.90.074005} {\bibfield  {journal}
  {\bibinfo  {journal} {Phys. Rev.}\ }\textbf {\bibinfo {volume} {D90}},\
  \bibinfo {pages} {074005} (\bibinfo {year} {2014})},\ \Eprint
  {http://arxiv.org/abs/1408.0692} {arXiv:1408.0692 [hep-ph]} \BibitemShut
  {NoStop}%
\bibitem [{\citenamefont {Troshin}\ and\ \citenamefont
  {Tyurin}(2016)}]{Troshin:2016frs}%
  \BibitemOpen
  \bibfield  {author} {\bibinfo {author} {\bibfnamefont {S.~M.}\ \bibnamefont
  {Troshin}}\ and\ \bibinfo {author} {\bibfnamefont {N.~E.}\ \bibnamefont
  {Tyurin}},\ }\href {\doibase 10.1142/S0217732316500796} {\bibfield  {journal}
  {\bibinfo  {journal} {Mod. Phys. Lett.}\ }\textbf {\bibinfo {volume} {A31}},\
  \bibinfo {pages} {1650079} (\bibinfo {year} {2016})},\ \Eprint
  {http://arxiv.org/abs/1602.08972} {arXiv:1602.08972 [hep-ph]} \BibitemShut
  {NoStop}%
\bibitem [{\citenamefont {Troshin}\ and\ \citenamefont
  {Tyurin}(2017{\natexlab{a}})}]{Troshin:2017zmg}%
  \BibitemOpen
  \bibfield  {author} {\bibinfo {author} {\bibfnamefont {S.~M.}\ \bibnamefont
  {Troshin}}\ and\ \bibinfo {author} {\bibfnamefont {N.~E.}\ \bibnamefont
  {Tyurin}},\ }\href {\doibase 10.1140/epja/i2017-12246-1} {\bibfield
  {journal} {\bibinfo  {journal} {Eur. Phys. J.}\ }\textbf {\bibinfo {volume}
  {A53}},\ \bibinfo {pages} {57} (\bibinfo {year} {2017}{\natexlab{a}})},\
  \Eprint {http://arxiv.org/abs/1701.01815} {arXiv:1701.01815 [hep-ph]}
  \BibitemShut {NoStop}%
\bibitem [{\citenamefont {Troshin}\ and\ \citenamefont
  {Tyurin}(2017{\natexlab{b}})}]{Troshin:2017ucy}%
  \BibitemOpen
  \bibfield  {author} {\bibinfo {author} {\bibfnamefont {S.~M.}\ \bibnamefont
  {Troshin}}\ and\ \bibinfo {author} {\bibfnamefont {N.~E.}\ \bibnamefont
  {Tyurin}},\ }\href {\doibase 10.1142/S0217751X17501032} {\bibfield  {journal}
  {\bibinfo  {journal} {Int. J. Mod. Phys.}\ }\textbf {\bibinfo {volume}
  {A32}},\ \bibinfo {pages} {1750103} (\bibinfo {year} {2017}{\natexlab{b}})},\
  \Eprint {http://arxiv.org/abs/1704.00443} {arXiv:1704.00443 [hep-ph]}
  \BibitemShut {NoStop}%
\bibitem [{\citenamefont {Puzikov}\ \emph {et~al.}(1957)\citenamefont
  {Puzikov}, \citenamefont {Ryndin},\ and\ \citenamefont
  {Smorodinsky}}]{puzikov1957construction}%
  \BibitemOpen
  \bibfield  {author} {\bibinfo {author} {\bibfnamefont {L.}~\bibnamefont
  {Puzikov}}, \bibinfo {author} {\bibfnamefont {R.}~\bibnamefont {Ryndin}}, \
  and\ \bibinfo {author} {\bibfnamefont {J.}~\bibnamefont {Smorodinsky}},\
  }\href@noop {} {\bibfield  {journal} {\bibinfo  {journal} {Nuclear Physics}\
  }\textbf {\bibinfo {volume} {3}},\ \bibinfo {pages} {436} (\bibinfo {year}
  {1957})}\BibitemShut {NoStop}%
\bibitem [{\citenamefont {Gribov}\ and\ \citenamefont
  {Volkov}(1963)}]{Gribov:1963gx}%
  \BibitemOpen
  \bibfield  {author} {\bibinfo {author} {\bibfnamefont {V.~N.}\ \bibnamefont
  {Gribov}}\ and\ \bibinfo {author} {\bibfnamefont {D.~V.}\ \bibnamefont
  {Volkov}},\ }\href@noop {} {\bibfield  {journal} {\bibinfo  {journal} {Sov.
  Phys. JETP}\ }\textbf {\bibinfo {volume} {17}},\ \bibinfo {pages} {720}
  (\bibinfo {year} {1963})},\ \bibinfo {note} {[Zh. Eksp. Teor.
  Fiz.44,1068(1963)]}\BibitemShut {NoStop}%
\bibitem [{\citenamefont {Sharp}\ and\ \citenamefont
  {Wagner}(1963)}]{Sharp:1963zz}%
  \BibitemOpen
  \bibfield  {author} {\bibinfo {author} {\bibfnamefont {D.~H.}\ \bibnamefont
  {Sharp}}\ and\ \bibinfo {author} {\bibfnamefont {W.~G.}\ \bibnamefont
  {Wagner}},\ }\href {\doibase 10.1103/PhysRev.131.2226} {\bibfield  {journal}
  {\bibinfo  {journal} {Phys. Rev.}\ }\textbf {\bibinfo {volume} {131}},\
  \bibinfo {pages} {2226} (\bibinfo {year} {1963})}\BibitemShut {NoStop}%
\bibitem [{\citenamefont {Selyugin}(2016)}]{Selyugin:2015vao}%
  \BibitemOpen
  \bibfield  {author} {\bibinfo {author} {\bibfnamefont {O.~V.}\ \bibnamefont
  {Selyugin}},\ }\href {\doibase 10.1134/S1547477116030195} {\bibfield
  {journal} {\bibinfo  {journal} {Phys. Part. Nucl. Lett.}\ }\textbf {\bibinfo
  {volume} {13}},\ \bibinfo {pages} {303} (\bibinfo {year} {2016})},\ \Eprint
  {http://arxiv.org/abs/1512.05130} {arXiv:1512.05130 [hep-ph]} \BibitemShut
  {NoStop}%
\bibitem [{\citenamefont {Gersten}(1969)}]{Gersten:1969ae}%
  \BibitemOpen
  \bibfield  {author} {\bibinfo {author} {\bibfnamefont {A.}~\bibnamefont
  {Gersten}},\ }\href {\doibase 10.1016/0550-3213(69)90072-8} {\bibfield
  {journal} {\bibinfo  {journal} {Nucl. Phys.}\ }\textbf {\bibinfo {volume}
  {B12}},\ \bibinfo {pages} {537} (\bibinfo {year} {1969})}\BibitemShut
  {NoStop}%
\bibitem [{\citenamefont {Bowcock}\ and\ \citenamefont
  {Burkhardt}(1975)}]{Bowcock:1976ax}%
  \BibitemOpen
  \bibfield  {author} {\bibinfo {author} {\bibfnamefont {J.~E.}\ \bibnamefont
  {Bowcock}}\ and\ \bibinfo {author} {\bibfnamefont {H.}~\bibnamefont
  {Burkhardt}},\ }\href {\doibase 10.1088/0034-4885/38/9/002} {\bibfield
  {journal} {\bibinfo  {journal} {Rept. Prog. Phys.}\ }\textbf {\bibinfo
  {volume} {38}},\ \bibinfo {pages} {1099} (\bibinfo {year}
  {1975})}\BibitemShut {NoStop}%
\bibitem [{\citenamefont {Broniowski}\ and\ \citenamefont
  {Ruiz~Arriola}(2017{\natexlab{b}})}]{Broniowski:2017rhz}%
  \BibitemOpen
  \bibfield  {author} {\bibinfo {author} {\bibfnamefont {W.}~\bibnamefont
  {Broniowski}}\ and\ \bibinfo {author} {\bibfnamefont {E.}~\bibnamefont
  {Ruiz~Arriola}},\ }\bibfield  {booktitle} {\emph {\bibinfo {booktitle}
  {{Proceedings, 9th Workshop "Excited QCD" 2017: Sintra, Portugal, May 7-13,
  2017}}},\ }\href {\doibase 10.5506/APhysPolBSupp.10.1203} {\bibfield
  {journal} {\bibinfo  {journal} {Acta Phys. Polon. Supp.}\ }\textbf {\bibinfo
  {volume} {10}},\ \bibinfo {pages} {1203} (\bibinfo {year}
  {2017}{\natexlab{b}})},\ \Eprint {http://arxiv.org/abs/1708.00402}
  {arXiv:1708.00402 [nucl-th]} \BibitemShut {NoStop}%
\bibitem [{\citenamefont {Covolan}\ \emph {et~al.}(1996)\citenamefont
  {Covolan}, \citenamefont {Montanha},\ and\ \citenamefont
  {Goulianos}}]{Covolan:1996uy}%
  \BibitemOpen
  \bibfield  {author} {\bibinfo {author} {\bibfnamefont {R.~J.~M.}\
  \bibnamefont {Covolan}}, \bibinfo {author} {\bibfnamefont {J.}~\bibnamefont
  {Montanha}}, \ and\ \bibinfo {author} {\bibfnamefont {K.~A.}\ \bibnamefont
  {Goulianos}},\ }\href {\doibase 10.1016/S0370-2693(96)01362-7} {\bibfield
  {journal} {\bibinfo  {journal} {Phys. Lett.}\ }\textbf {\bibinfo {volume}
  {B389}},\ \bibinfo {pages} {176} (\bibinfo {year} {1996})}\BibitemShut
  {NoStop}%
\bibitem [{\citenamefont {Phillips}\ and\ \citenamefont
  {Barger}(1973)}]{Phillips:1974vt}%
  \BibitemOpen
  \bibfield  {author} {\bibinfo {author} {\bibfnamefont {R.}~\bibnamefont
  {Phillips}}\ and\ \bibinfo {author} {\bibfnamefont {V.~D.}\ \bibnamefont
  {Barger}},\ }\href {\doibase 10.1016/0370-2693(73)90154-8} {\bibfield
  {journal} {\bibinfo  {journal} {Phys.Lett.}\ }\textbf {\bibinfo {volume}
  {B46}},\ \bibinfo {pages} {412} (\bibinfo {year} {1973})}\BibitemShut
  {NoStop}%
\bibitem [{\citenamefont {Fagundes}\ \emph {et~al.}(2013)\citenamefont
  {Fagundes}, \citenamefont {Grau}, \citenamefont {Pacetti}, \citenamefont
  {Pancheri},\ and\ \citenamefont {Srivastava}}]{Fagundes:2013aja}%
  \BibitemOpen
  \bibfield  {author} {\bibinfo {author} {\bibfnamefont {D.~A.}\ \bibnamefont
  {Fagundes}}, \bibinfo {author} {\bibfnamefont {A.}~\bibnamefont {Grau}},
  \bibinfo {author} {\bibfnamefont {S.}~\bibnamefont {Pacetti}}, \bibinfo
  {author} {\bibfnamefont {G.}~\bibnamefont {Pancheri}}, \ and\ \bibinfo
  {author} {\bibfnamefont {Y.~N.}\ \bibnamefont {Srivastava}},\ }\href
  {\doibase 10.1103/PhysRevD.88.094019} {\bibfield  {journal} {\bibinfo
  {journal} {Phys.Rev.}\ }\textbf {\bibinfo {volume} {D88}},\ \bibinfo {pages}
  {094019} (\bibinfo {year} {2013})},\ \Eprint {http://arxiv.org/abs/1306.0452}
  {arXiv:1306.0452 [hep-ph]} \BibitemShut {NoStop}%
\bibitem [{\citenamefont {Bell}\ and\ \citenamefont
  {Goebel}(1965)}]{Bell:1964fz}%
  \BibitemOpen
  \bibfield  {author} {\bibinfo {author} {\bibfnamefont {J.~S.}\ \bibnamefont
  {Bell}}\ and\ \bibinfo {author} {\bibfnamefont {C.~J.}\ \bibnamefont
  {Goebel}},\ }\href {\doibase 10.1103/PhysRev.138.B1198} {\bibfield  {journal}
  {\bibinfo  {journal} {Phys. Rev.}\ }\textbf {\bibinfo {volume} {138}},\
  \bibinfo {pages} {B1198} (\bibinfo {year} {1965})}\BibitemShut {NoStop}%
\bibitem [{\citenamefont {Bia\l{}kowski}(1970)}]{bialkowski1970phenomenology}%
  \BibitemOpen
  \bibfield  {author} {\bibinfo {author} {\bibfnamefont {G.}~\bibnamefont
  {Bia\l{}kowski}},\ }\href@noop {} {\bibfield  {journal} {\bibinfo  {journal}
  {Acta Phys. Pol. B}\ }\textbf {\bibinfo {volume} {1}},\ \bibinfo {pages}
  {109} (\bibinfo {year} {1970})}\BibitemShut {NoStop}%
\bibitem [{\citenamefont {Jenkovszky}\ and\ \citenamefont
  {Paccanoni}(1981)}]{Jenkovszky:1981xv}%
  \BibitemOpen
  \bibfield  {author} {\bibinfo {author} {\bibfnamefont {L.~L.}\ \bibnamefont
  {Jenkovszky}}\ and\ \bibinfo {author} {\bibfnamefont {F.}~\bibnamefont
  {Paccanoni}},\ }\href {\doibase 10.1007/BF02904446} {\bibfield  {journal}
  {\bibinfo  {journal} {Nuovo Cim.}\ }\textbf {\bibinfo {volume} {A62}},\
  \bibinfo {pages} {133} (\bibinfo {year} {1981})}\BibitemShut {NoStop}%
\bibitem [{\citenamefont {Jenkovszky}\ and\ \citenamefont
  {Wall}(1976)}]{Jenkovszky:1976sf}%
  \BibitemOpen
  \bibfield  {author} {\bibinfo {author} {\bibfnamefont {L.~L.}\ \bibnamefont
  {Jenkovszky}}\ and\ \bibinfo {author} {\bibfnamefont {A.~N.}\ \bibnamefont
  {Wall}},\ }\href {\doibase 10.1007/BF01587265} {\bibfield  {journal}
  {\bibinfo  {journal} {Czech. J. Phys.}\ }\textbf {\bibinfo {volume} {B26}},\
  \bibinfo {pages} {447} (\bibinfo {year} {1976})}\BibitemShut {NoStop}%
\bibitem [{\citenamefont {Saleem}\ and\ \citenamefont
  {{Fazal-e-Aleem}}(1981)}]{Saleem:1981av}%
  \BibitemOpen
  \bibfield  {author} {\bibinfo {author} {\bibfnamefont {M.}~\bibnamefont
  {Saleem}}\ and\ \bibinfo {author} {\bibnamefont {{Fazal-e-Aleem}}},\
  }\href@noop {} {\bibfield  {journal} {\bibinfo  {journal} {Hadronic J.}\
  }\textbf {\bibinfo {volume} {5}},\ \bibinfo {pages} {71} (\bibinfo {year}
  {1981})}\BibitemShut {NoStop}%
\bibitem [{\citenamefont {Jenkovszky}\ \emph {et~al.}(1987)\citenamefont
  {Jenkovszky}, \citenamefont {Struminsky},\ and\ \citenamefont
  {Shelkovenko}}]{Jenkovszky:1987gv}%
  \BibitemOpen
  \bibfield  {author} {\bibinfo {author} {\bibfnamefont {L.~L.}\ \bibnamefont
  {Jenkovszky}}, \bibinfo {author} {\bibfnamefont {B.~V.}\ \bibnamefont
  {Struminsky}}, \ and\ \bibinfo {author} {\bibfnamefont {A.~N.}\ \bibnamefont
  {Shelkovenko}},\ }\href {\doibase 10.1007/BF01573947} {\bibfield  {journal}
  {\bibinfo  {journal} {Z. Phys.}\ }\textbf {\bibinfo {volume} {C36}},\
  \bibinfo {pages} {495} (\bibinfo {year} {1987})}\BibitemShut {NoStop}%
\bibitem [{\citenamefont {Jenkovszky}\ \emph {et~al.}(1990)\citenamefont
  {Jenkovszky}, \citenamefont {Martynov},\ and\ \citenamefont
  {Struminsky}}]{Jenkovszky:1990ki}%
  \BibitemOpen
  \bibfield  {author} {\bibinfo {author} {\bibfnamefont {L.~L.}\ \bibnamefont
  {Jenkovszky}}, \bibinfo {author} {\bibfnamefont {E.~S.}\ \bibnamefont
  {Martynov}}, \ and\ \bibinfo {author} {\bibfnamefont {B.~V.}\ \bibnamefont
  {Struminsky}},\ }\href {\doibase 10.1016/0370-2693(90)91031-6} {\bibfield
  {journal} {\bibinfo  {journal} {Phys. Lett.}\ }\textbf {\bibinfo {volume}
  {B249}},\ \bibinfo {pages} {535} (\bibinfo {year} {1990})}\BibitemShut
  {NoStop}%
\bibitem [{\citenamefont {Jenkovszky}\ \emph {et~al.}(2011)\citenamefont
  {Jenkovszky}, \citenamefont {Lengyel},\ and\ \citenamefont
  {Lontkovskyi}}]{Jenkovszky:2011hu}%
  \BibitemOpen
  \bibfield  {author} {\bibinfo {author} {\bibfnamefont {L.~L.}\ \bibnamefont
  {Jenkovszky}}, \bibinfo {author} {\bibfnamefont {A.~I.}\ \bibnamefont
  {Lengyel}}, \ and\ \bibinfo {author} {\bibfnamefont {D.~I.}\ \bibnamefont
  {Lontkovskyi}},\ }\href {\doibase 10.1142/S0217751X11054760} {\bibfield
  {journal} {\bibinfo  {journal} {Int. J. Mod. Phys.}\ }\textbf {\bibinfo
  {volume} {A26}},\ \bibinfo {pages} {4755} (\bibinfo {year} {2011})},\ \Eprint
  {http://arxiv.org/abs/1105.1202} {arXiv:1105.1202 [hep-ph]} \BibitemShut
  {NoStop}%
\bibitem [{\citenamefont {Martynov}(2007)}]{Martynov:2007dy}%
  \BibitemOpen
  \bibfield  {author} {\bibinfo {author} {\bibfnamefont {E.}~\bibnamefont
  {Martynov}},\ }\href {\doibase 10.1103/PhysRevD.76.074030} {\bibfield
  {journal} {\bibinfo  {journal} {Phys. Rev.}\ }\textbf {\bibinfo {volume}
  {D76}},\ \bibinfo {pages} {074030} (\bibinfo {year} {2007})},\ \Eprint
  {http://arxiv.org/abs/hep-ph/0703248} {arXiv:hep-ph/0703248 [HEP-PH]}
  \BibitemShut {NoStop}%
\bibitem [{\citenamefont {Alkin}\ \emph {et~al.}(2012)\citenamefont {Alkin},
  \citenamefont {Cudell},\ and\ \citenamefont {Martynov}}]{Alkin:2011if}%
  \BibitemOpen
  \bibfield  {author} {\bibinfo {author} {\bibfnamefont {A.}~\bibnamefont
  {Alkin}}, \bibinfo {author} {\bibfnamefont {J.~R.}\ \bibnamefont {Cudell}}, \
  and\ \bibinfo {author} {\bibfnamefont {E.}~\bibnamefont {Martynov}},\
  }\bibfield  {booktitle} {\emph {\bibinfo {booktitle} {{Proceedings, 30 Years
  of Strong Interactions: A three-day Meeting in honor of Joseph Cugnon and
  Hans-Jürgen Pirner: Spa, Liege, Belgium, April 6-8, 2011}}},\ }\href
  {\doibase 10.1007/s00601-012-0306-5} {\bibfield  {journal} {\bibinfo
  {journal} {Few Body Syst.}\ }\textbf {\bibinfo {volume} {53}},\ \bibinfo
  {pages} {87} (\bibinfo {year} {2012})},\ \Eprint
  {http://arxiv.org/abs/1109.1306} {arXiv:1109.1306 [hep-ph]} \BibitemShut
  {NoStop}%
\bibitem [{\citenamefont {Martynov}\ and\ \citenamefont
  {Nicolescu}(2018{\natexlab{a}})}]{Martynov:2017zjz}%
  \BibitemOpen
  \bibfield  {author} {\bibinfo {author} {\bibfnamefont {E.}~\bibnamefont
  {Martynov}}\ and\ \bibinfo {author} {\bibfnamefont {B.}~\bibnamefont
  {Nicolescu}},\ }\href {\doibase 10.1016/j.physletb.2018.01.054} {\bibfield
  {journal} {\bibinfo  {journal} {Phys. Lett.}\ }\textbf {\bibinfo {volume}
  {B778}},\ \bibinfo {pages} {414} (\bibinfo {year} {2018}{\natexlab{a}})},\
  \Eprint {http://arxiv.org/abs/1711.03288} {arXiv:1711.03288 [hep-ph]}
  \BibitemShut {NoStop}%
\bibitem [{\citenamefont {Khoze}\ \emph
  {et~al.}(2018{\natexlab{a}})\citenamefont {Khoze}, \citenamefont {Martin},\
  and\ \citenamefont {Ryskin}}]{Khoze:2017swe}%
  \BibitemOpen
  \bibfield  {author} {\bibinfo {author} {\bibfnamefont {V.~A.}\ \bibnamefont
  {Khoze}}, \bibinfo {author} {\bibfnamefont {A.~D.}\ \bibnamefont {Martin}}, \
  and\ \bibinfo {author} {\bibfnamefont {M.~G.}\ \bibnamefont {Ryskin}},\
  }\href {\doibase 10.1103/PhysRevD.97.034019} {\bibfield  {journal} {\bibinfo
  {journal} {Phys. Rev.}\ }\textbf {\bibinfo {volume} {D97}},\ \bibinfo {pages}
  {034019} (\bibinfo {year} {2018}{\natexlab{a}})},\ \Eprint
  {http://arxiv.org/abs/1712.00325} {arXiv:1712.00325 [hep-ph]} \BibitemShut
  {NoStop}%
\bibitem [{\citenamefont {Khoze}\ \emph
  {et~al.}(2018{\natexlab{b}})\citenamefont {Khoze}, \citenamefont {Martin},\
  and\ \citenamefont {Ryskin}}]{Khoze:2018bus}%
  \BibitemOpen
  \bibfield  {author} {\bibinfo {author} {\bibfnamefont {V.~A.}\ \bibnamefont
  {Khoze}}, \bibinfo {author} {\bibfnamefont {A.~D.}\ \bibnamefont {Martin}}, \
  and\ \bibinfo {author} {\bibfnamefont {M.~G.}\ \bibnamefont {Ryskin}},\
  }\href {\doibase 10.1016/j.physletb.2018.03.025} {\bibfield  {journal}
  {\bibinfo  {journal} {Phys. Lett.}\ }\textbf {\bibinfo {volume} {B780}},\
  \bibinfo {pages} {352} (\bibinfo {year} {2018}{\natexlab{b}})},\ \Eprint
  {http://arxiv.org/abs/1801.07065} {arXiv:1801.07065 [hep-ph]} \BibitemShut
  {NoStop}%
\bibitem [{\citenamefont {Martynov}\ and\ \citenamefont
  {Nicolescu}(2018{\natexlab{b}})}]{Martynov:2018nyb}%
  \BibitemOpen
  \bibfield  {author} {\bibinfo {author} {\bibfnamefont {E.}~\bibnamefont
  {Martynov}}\ and\ \bibinfo {author} {\bibfnamefont {B.}~\bibnamefont
  {Nicolescu}},\ }\href {\doibase 10.1016/j.physletb.2018.09.049} {\bibfield
  {journal} {\bibinfo  {journal} {Phys. Lett.}\ }\textbf {\bibinfo {volume}
  {B786}},\ \bibinfo {pages} {207} (\bibinfo {year} {2018}{\natexlab{b}})},\
  \Eprint {http://arxiv.org/abs/1804.10139} {arXiv:1804.10139 [hep-ph]}
  \BibitemShut {NoStop}%
\bibitem [{\citenamefont {Kontros}\ \emph {et~al.}(2000)\citenamefont
  {Kontros}, \citenamefont {Kontros},\ and\ \citenamefont {Lengyel}}]{KKL1}%
  \BibitemOpen
  \bibfield  {author} {\bibinfo {author} {\bibfnamefont {J.}~\bibnamefont
  {Kontros}}, \bibinfo {author} {\bibfnamefont {K.}~\bibnamefont {Kontros}}, \
  and\ \bibinfo {author} {\bibfnamefont {A.}~\bibnamefont {Lengyel}},\ }in\
  \href@noop {} {\emph {\bibinfo {booktitle} {{New trends in high-energy
  physics: Experiment, phenomenology, theory. Proceedings, International
  School-Conference, Crimea 2000, Yalta, Ukraine, May 27-June 4, 2000}}}}\
  (\bibinfo {year} {2000})\ pp.\ \bibinfo {pages} {140--144},\ \Eprint
  {http://arxiv.org/abs/hep-ph/0006141} {arXiv:hep-ph/0006141 [hep-ph]}
  \BibitemShut {NoStop}%
\bibitem [{\citenamefont {Kontros}\ \emph {et~al.}(2001)\citenamefont
  {Kontros}, \citenamefont {Kontros},\ and\ \citenamefont {Lengyel}}]{KKL}%
  \BibitemOpen
  \bibfield  {author} {\bibinfo {author} {\bibfnamefont {J.}~\bibnamefont
  {Kontros}}, \bibinfo {author} {\bibfnamefont {K.}~\bibnamefont {Kontros}}, \
  and\ \bibinfo {author} {\bibfnamefont {A.}~\bibnamefont {Lengyel}},\ }in\
  \href@noop {} {\emph {\bibinfo {booktitle} {{Elastic and diffractive
  scattering. Proceedings, 9th Blois Workshop, Pruhonice, Czech Republic, June
  9-15, 2001}}}}\ (\bibinfo {year} {2001})\ pp.\ \bibinfo {pages} {287--292},\
  \Eprint {http://arxiv.org/abs/hep-ph/0104133} {arXiv:hep-ph/0104133 [hep-ph]}
  \BibitemShut {NoStop}%
\bibitem [{\citenamefont {Antchev}\ \emph
  {et~al.}(2013{\natexlab{a}})\citenamefont {Antchev} \emph {et~al.}}]{totem7}%
  \BibitemOpen
  \bibfield  {author} {\bibinfo {author} {\bibfnamefont {G.}~\bibnamefont
  {Antchev}} \emph {et~al.} (\bibinfo {collaboration} {TOTEM}),\ }\href
  {\doibase 10.1209/0295-5075/101/21002} {\bibfield  {journal} {\bibinfo
  {journal} {EPL}\ }\textbf {\bibinfo {volume} {101}},\ \bibinfo {pages}
  {21002} (\bibinfo {year} {2013}{\natexlab{a}})}\BibitemShut {NoStop}%
\bibitem [{bar()}]{barppdcs}%
  \BibitemOpen
  \href@noop {} {}\bibinfo {howpublished} {\url{
  http://durpdg.dur.ac.uk}}\BibitemShut {NoStop}%
\bibitem [{\citenamefont {Abreu}\ \emph {et~al.}(2012)\citenamefont {Abreu}
  \emph {et~al.}}]{Auger}%
  \BibitemOpen
  \bibfield  {author} {\bibinfo {author} {\bibfnamefont {P.}~\bibnamefont
  {Abreu}} \emph {et~al.} (\bibinfo {collaboration} {Pierre Auger}),\ }\href
  {\doibase 10.1103/PhysRevLett.109.062002} {\bibfield  {journal} {\bibinfo
  {journal} {Phys. Rev. Lett.}\ }\textbf {\bibinfo {volume} {109}},\ \bibinfo
  {pages} {062002} (\bibinfo {year} {2012})},\ \Eprint
  {http://arxiv.org/abs/1208.1520} {arXiv:1208.1520 [hep-ex]} \BibitemShut
  {NoStop}%
\bibitem [{\citenamefont {Antchev}\ \emph
  {et~al.}(2013{\natexlab{b}})\citenamefont {Antchev} \emph
  {et~al.}}]{totem81}%
  \BibitemOpen
  \bibfield  {author} {\bibinfo {author} {\bibfnamefont {G.}~\bibnamefont
  {Antchev}} \emph {et~al.} (\bibinfo {collaboration} {TOTEM}),\ }\href
  {\doibase 10.1103/PhysRevLett.111.012001} {\bibfield  {journal} {\bibinfo
  {journal} {Phys. Rev. Lett.}\ }\textbf {\bibinfo {volume} {111}},\ \bibinfo
  {pages} {012001} (\bibinfo {year} {2013}{\natexlab{b}})}\BibitemShut
  {NoStop}%
\bibitem [{\citenamefont {Antchev}\ \emph
  {et~al.}(2013{\natexlab{c}})\citenamefont {Antchev} \emph
  {et~al.}}]{totem72}%
  \BibitemOpen
  \bibfield  {author} {\bibinfo {author} {\bibfnamefont {G.}~\bibnamefont
  {Antchev}} \emph {et~al.} (\bibinfo {collaboration} {TOTEM}),\ }\href
  {\doibase 10.1209/0295-5075/101/21004} {\bibfield  {journal} {\bibinfo
  {journal} {EPL}\ }\textbf {\bibinfo {volume} {101}},\ \bibinfo {pages}
  {21004} (\bibinfo {year} {2013}{\natexlab{c}})}\BibitemShut {NoStop}%
\bibitem [{\citenamefont {Patrignani}\ \emph {et~al.}(2016)\citenamefont
  {Patrignani} \emph {et~al.}}]{PDG}%
  \BibitemOpen
  \bibfield  {author} {\bibinfo {author} {\bibfnamefont {C.}~\bibnamefont
  {Patrignani}} \emph {et~al.} (\bibinfo {collaboration} {Particle Data
  Group}),\ }\href {\doibase 10.1088/1674-1137/40/10/100001} {\bibfield
  {journal} {\bibinfo  {journal} {Chin. Phys.}\ }\textbf {\bibinfo {volume}
  {C40}},\ \bibinfo {pages} {100001} (\bibinfo {year} {2016})}\BibitemShut
  {NoStop}%
\bibitem [{\citenamefont {Antchev}\ \emph
  {et~al.}(2017{\natexlab{a}})\citenamefont {Antchev} \emph {et~al.}}]{Giani}%
  \BibitemOpen
  \bibfield  {author} {\bibinfo {author} {\bibfnamefont {G.}~\bibnamefont
  {Antchev}} \emph {et~al.} (\bibinfo {collaboration} {TOTEM}),\ }\href@noop {}
  {\  (\bibinfo {year} {2017}{\natexlab{a}})},\ \Eprint
  {http://arxiv.org/abs/1712.06153} {arXiv:1712.06153 [hep-ex]} \BibitemShut
  {NoStop}%
\bibitem [{\citenamefont {Antchev}\ \emph {et~al.}(2016)\citenamefont {Antchev}
  \emph {et~al.}}]{totem82}%
  \BibitemOpen
  \bibfield  {author} {\bibinfo {author} {\bibfnamefont {G.}~\bibnamefont
  {Antchev}} \emph {et~al.} (\bibinfo {collaboration} {TOTEM}),\ }\href
  {\doibase 10.1140/epjc/s10052-016-4399-8} {\bibfield  {journal} {\bibinfo
  {journal} {Eur. Phys. J.}\ }\textbf {\bibinfo {volume} {C76}},\ \bibinfo
  {pages} {661} (\bibinfo {year} {2016})},\ \Eprint
  {http://arxiv.org/abs/1610.00603} {arXiv:1610.00603 [nucl-ex]} \BibitemShut
  {NoStop}%
\bibitem [{\citenamefont {Antchev}\ \emph
  {et~al.}(2017{\natexlab{b}})\citenamefont {Antchev} \emph
  {et~al.}}]{TOTEM_rho}%
  \BibitemOpen
  \bibfield  {author} {\bibinfo {author} {\bibfnamefont {G.}~\bibnamefont
  {Antchev}} \emph {et~al.} (\bibinfo {collaboration} {TOTEM}),\ }\href@noop {}
  {\  (\bibinfo {year} {2017}{\natexlab{b}})},\ \bibinfo {note}
  {{CERN-EP-2017-335}}\BibitemShut {NoStop}%
\bibitem [{\citenamefont {Wall}\ \emph {et~al.}(1988)\citenamefont {Wall},
  \citenamefont {Jenkovszky},\ and\ \citenamefont {Struminsky}}]{PEPAN}%
  \BibitemOpen
  \bibfield  {author} {\bibinfo {author} {\bibfnamefont {A.}~\bibnamefont
  {Wall}}, \bibinfo {author} {\bibfnamefont {L.}~\bibnamefont {Jenkovszky}}, \
  and\ \bibinfo {author} {\bibfnamefont {B.}~\bibnamefont {Struminsky}},\
  }\href@noop {} {\bibfield  {journal} {\bibinfo  {journal} {Sov. J. Particles
  and Nuclei}\ }\textbf {\bibinfo {volume} {19}},\ \bibinfo {pages} {180}
  (\bibinfo {year} {1988})}\BibitemShut {NoStop}%
\bibitem [{\citenamefont {Kundrat}\ \emph {et~al.}(1981)\citenamefont
  {Kundrat}, \citenamefont {Lokajicek},\ and\ \citenamefont
  {Lokajicek}}]{Kundrat:1980es}%
  \BibitemOpen
  \bibfield  {author} {\bibinfo {author} {\bibfnamefont {V.}~\bibnamefont
  {Kundrat}}, \bibinfo {author} {\bibfnamefont {M.}~\bibnamefont {Lokajicek}},
  \ and\ \bibinfo {author} {\bibfnamefont {M.~V.}\ \bibnamefont {Lokajicek}},\
  }\href {\doibase 10.1007/BF01595377} {\bibfield  {journal} {\bibinfo
  {journal} {Czech. J. Phys.}\ }\textbf {\bibinfo {volume} {B31}},\ \bibinfo
  {pages} {1334} (\bibinfo {year} {1981})}\BibitemShut {NoStop}%
\bibitem [{\citenamefont {Kundrat}\ and\ \citenamefont
  {Lokajicek}(1994)}]{Kundrat:1993sv}%
  \BibitemOpen
  \bibfield  {author} {\bibinfo {author} {\bibfnamefont {V.}~\bibnamefont
  {Kundrat}}\ and\ \bibinfo {author} {\bibfnamefont {M.}~\bibnamefont
  {Lokajicek}},\ }\href {\doibase 10.1007/BF01557628} {\bibfield  {journal}
  {\bibinfo  {journal} {Z. Phys.}\ }\textbf {\bibinfo {volume} {C63}},\
  \bibinfo {pages} {619} (\bibinfo {year} {1994})}\BibitemShut {NoStop}%
\bibitem [{\citenamefont {Prochazka}\ and\ \citenamefont
  {Kundrat}(2016)}]{Prochazka:2016wno}%
  \BibitemOpen
  \bibfield  {author} {\bibinfo {author} {\bibfnamefont {J.}~\bibnamefont
  {Prochazka}}\ and\ \bibinfo {author} {\bibfnamefont {V.}~\bibnamefont
  {Kundrat}},\ }\href@noop {} {\  (\bibinfo {year} {2016})},\ \Eprint
  {http://arxiv.org/abs/1606.09479} {arXiv:1606.09479 [hep-th]} \BibitemShut
  {NoStop}%
\bibitem [{\citenamefont {Prochazka}\ \emph {et~al.}(2017)\citenamefont
  {Prochazka}, \citenamefont {Kundrat},\ and\ \citenamefont
  {Lokajicek}}]{Prochazka:2017tby}%
  \BibitemOpen
  \bibfield  {author} {\bibinfo {author} {\bibfnamefont {J.}~\bibnamefont
  {Prochazka}}, \bibinfo {author} {\bibfnamefont {V.}~\bibnamefont {Kundrat}},
  \ and\ \bibinfo {author} {\bibfnamefont {M.~V.}\ \bibnamefont {Lokajicek}},\
  }in\ \href {http://inspirehep.net/record/1633451/files/arXiv:1710.10640.pdf}
  {\emph {\bibinfo {booktitle} {{17th conference on Elastic and Diffractive
  Scattering (EDS 17) Prague, Czech Republic, June 26-30, 2017}}}}\ (\bibinfo
  {year} {2017})\ \Eprint {http://arxiv.org/abs/1710.10640} {arXiv:1710.10640
  [hep-ph]} \BibitemShut {NoStop}%
\bibitem [{\citenamefont {Petrov}(2018)}]{Petrov:2018xma}%
  \BibitemOpen
  \bibfield  {author} {\bibinfo {author} {\bibfnamefont {V.~A.}\ \bibnamefont
  {Petrov}},\ }\href {\doibase 10.1140/epjc/s10052-018-5716-1,
  10.1140/epjc/s10052-018-5889-7} {\bibfield  {journal} {\bibinfo  {journal}
  {Eur. Phys. J.}\ }\textbf {\bibinfo {volume} {C78}},\ \bibinfo {pages} {221}
  (\bibinfo {year} {2018})},\ \bibinfo {note} {[Erratum: Eur. Phys.
  J.C78,no.5,414(2018)]},\ \Eprint {http://arxiv.org/abs/1801.01815}
  {arXiv:1801.01815 [hep-ph]} \BibitemShut {NoStop}%
\bibitem [{\citenamefont {Troshin}\ and\ \citenamefont
  {Tyurin}(1999)}]{Troshin:1998ij}%
  \BibitemOpen
  \bibfield  {author} {\bibinfo {author} {\bibfnamefont {S.~M.}\ \bibnamefont
  {Troshin}}\ and\ \bibinfo {author} {\bibfnamefont {N.~E.}\ \bibnamefont
  {Tyurin}},\ }\href {\doibase 10.1134/1.953118} {\bibfield  {journal}
  {\bibinfo  {journal} {Phys. Part. Nucl.}\ }\textbf {\bibinfo {volume} {30}},\
  \bibinfo {pages} {550} (\bibinfo {year} {1999})},\ \Eprint
  {http://arxiv.org/abs/hep-ph/9810495} {arXiv:hep-ph/9810495 [hep-ph]}
  \BibitemShut {NoStop}%
\bibitem [{\citenamefont {Cudell}\ \emph {et~al.}(2009)\citenamefont {Cudell},
  \citenamefont {Predazzi},\ and\ \citenamefont {Selyugin}}]{Cudell:2008yb}%
  \BibitemOpen
  \bibfield  {author} {\bibinfo {author} {\bibfnamefont {J.~R.}\ \bibnamefont
  {Cudell}}, \bibinfo {author} {\bibfnamefont {E.}~\bibnamefont {Predazzi}}, \
  and\ \bibinfo {author} {\bibfnamefont {O.~V.}\ \bibnamefont {Selyugin}},\
  }\href {\doibase 10.1103/PhysRevD.79.034033} {\bibfield  {journal} {\bibinfo
  {journal} {Phys. Rev.}\ }\textbf {\bibinfo {volume} {D79}},\ \bibinfo {pages}
  {034033} (\bibinfo {year} {2009})},\ \Eprint {http://arxiv.org/abs/0812.0735}
  {arXiv:0812.0735 [hep-ph]} \BibitemShut {NoStop}%
\bibitem [{\citenamefont {Blankenbecler}\ and\ \citenamefont
  {Goldberger}(1962)}]{Blankenbecler:1962ez}%
  \BibitemOpen
  \bibfield  {author} {\bibinfo {author} {\bibfnamefont {R.}~\bibnamefont
  {Blankenbecler}}\ and\ \bibinfo {author} {\bibfnamefont {M.}~\bibnamefont
  {Goldberger}},\ }\href {\doibase 10.1103/PhysRev.126.766} {\bibfield
  {journal} {\bibinfo  {journal} {Phys.Rev.}\ }\textbf {\bibinfo {volume}
  {126}},\ \bibinfo {pages} {766} (\bibinfo {year} {1962})}\BibitemShut
  {NoStop}%
\bibitem [{\citenamefont {Pereira~da Silva}\ \emph {et~al.}(2007)\citenamefont
  {Pereira~da Silva}, \citenamefont {Menon},\ and\ \citenamefont
  {Avila}}]{daSilva:2008rv}%
  \BibitemOpen
  \bibfield  {author} {\bibinfo {author} {\bibfnamefont {G.~L.}\ \bibnamefont
  {Pereira~da Silva}}, \bibinfo {author} {\bibfnamefont {M.~J.}\ \bibnamefont
  {Menon}}, \ and\ \bibinfo {author} {\bibfnamefont {R.~F.}\ \bibnamefont
  {Avila}},\ }\bibfield  {booktitle} {\emph {\bibinfo {booktitle} {{Hadron
  physics. Proceedings, 10th International Workshop, Florianopolis, Brazil,
  April 26-31, 2007}}},\ }\href {\doibase 10.1142/S0218301307008732} {\bibfield
   {journal} {\bibinfo  {journal} {Int. J. Mod. Phys.}\ }\textbf {\bibinfo
  {volume} {E16}},\ \bibinfo {pages} {2923} (\bibinfo {year} {2007})},\ \Eprint
  {http://arxiv.org/abs/0802.1686} {arXiv:0802.1686 [hep-ph]} \BibitemShut
  {NoStop}%
\bibitem [{\citenamefont {Fagundes}\ \emph {et~al.}(2011)\citenamefont
  {Fagundes}, \citenamefont {Menon},\ and\ \citenamefont
  {Silva}}]{Fagundes:2011jh}%
  \BibitemOpen
  \bibfield  {author} {\bibinfo {author} {\bibfnamefont {D.~A.}\ \bibnamefont
  {Fagundes}}, \bibinfo {author} {\bibfnamefont {M.~J.}\ \bibnamefont {Menon}},
  \ and\ \bibinfo {author} {\bibfnamefont {G.~L.~P.}\ \bibnamefont {Silva}},\
  }\href {\doibase 10.1140/epjc/s10052-011-1637-y} {\bibfield  {journal}
  {\bibinfo  {journal} {Eur. Phys. J.}\ }\textbf {\bibinfo {volume} {C71}},\
  \bibinfo {pages} {1637} (\bibinfo {year} {2011})},\ \Eprint
  {http://arxiv.org/abs/1102.5028} {arXiv:1102.5028 [hep-ph]} \BibitemShut
  {NoStop}%
\bibitem [{\citenamefont {Selyugin}(2015)}]{Selyugin:2015pha}%
  \BibitemOpen
  \bibfield  {author} {\bibinfo {author} {\bibfnamefont {O.~V.}\ \bibnamefont
  {Selyugin}},\ }\href {\doibase 10.1103/PhysRevD.91.113003,
  10.1103/PhysRevD.92.099901} {\bibfield  {journal} {\bibinfo  {journal} {Phys.
  Rev.}\ }\textbf {\bibinfo {volume} {D91}},\ \bibinfo {pages} {113003}
  (\bibinfo {year} {2015})},\ \bibinfo {note} {[Erratum: Phys.
  Rev.D92,no.9,099901(2015)]},\ \Eprint {http://arxiv.org/abs/1505.02426}
  {arXiv:1505.02426 [hep-ph]} \BibitemShut {NoStop}%
\bibitem [{\citenamefont {Petrov}\ and\ \citenamefont
  {Samokhin}(2018)}]{Petrov:2018wlv}%
  \BibitemOpen
  \bibfield  {author} {\bibinfo {author} {\bibfnamefont {V.~A.}\ \bibnamefont
  {Petrov}}\ and\ \bibinfo {author} {\bibfnamefont {A.~P.}\ \bibnamefont
  {Samokhin}},\ }in\ \href
  {http://inspirehep.net/record/1647568/files/arXiv:1801.03809.pdf} {\emph
  {\bibinfo {booktitle} {{31th International Workshop on High Energy Physics:
  Critical points in the modern particle physics (IHEP2017) Protvino, Russia,
  July 5-7, 2017}}}}\ (\bibinfo {year} {2018})\ \Eprint
  {http://arxiv.org/abs/1801.03809} {arXiv:1801.03809 [hep-ph]} \BibitemShut
  {NoStop}%
\bibitem [{\citenamefont {Alkin}\ \emph {et~al.}(2018)\citenamefont {Alkin},
  \citenamefont {Martynov}, \citenamefont {Kovalenko},\ and\ \citenamefont
  {Troshin}}]{Alkin:2018vhh}%
  \BibitemOpen
  \bibfield  {author} {\bibinfo {author} {\bibfnamefont {A.}~\bibnamefont
  {Alkin}}, \bibinfo {author} {\bibfnamefont {E.}~\bibnamefont {Martynov}},
  \bibinfo {author} {\bibfnamefont {O.}~\bibnamefont {Kovalenko}}, \ and\
  \bibinfo {author} {\bibfnamefont {S.~M.}\ \bibnamefont {Troshin}},\
  }\href@noop {} {\  (\bibinfo {year} {2018})},\ \Eprint
  {http://arxiv.org/abs/1807.06471} {arXiv:1807.06471 [hep-ph]} \BibitemShut
  {NoStop}%
\end{thebibliography}%

\end{document}